\documentclass[pre,showpacs,preprintnumbers,amsmath,amssymb,twocolumn,nofootinbib,longbibliography]{revtex4-2}
\usepackage{multirow}
\usepackage{bm}
\usepackage{array}

\usepackage{graphicx}  
\usepackage{dcolumn}   
\usepackage{bm}        
\usepackage{amssymb}   
\usepackage{amsmath}
\usepackage{xspace}
\usepackage{xcolor}
\usepackage{float}

\usepackage [T1]{fontenc}
\usepackage [utf8]{inputenc}
\usepackage[colorlinks]{hyperref}

\usepackage[paperwidth=210mm,paperheight=297mm,centering,hmargin=2cm,vmargin=2.5cm]{geometry}
\usepackage{multirow}

\newcommand{\tw}{\ensuremath{t_\mathrm{w}}\xspace}
\newcommand{\teff}{\ensuremath{t^\mathrm{eff}}\xspace}

\newcommand{\Tm}{\ensuremath{T_\mathrm{m}}\xspace}

\newcommand{\Tg}{\ensuremath{T_\mathrm{g}}\xspace}

\begin{document}

\title{On the superposition principle and non-linear response in spin glasses}
\author{I.~Paga$^{1,*}$, Q.~Zhai$^{2,*}$, M.~Baity-Jesi$^3$, E.~Calore$^4$, A.~Cruz$^{5,6}$, C.~Cummings$^{7}$, L.~A.~Fernandez$^{8,6}$,  J.~M.~Gil-Narvion$^6$, I.~Gonzalez-Adalid~Pemartin$^8$, A.~Gordillo-Guerrero$^{9,10,6}$, D.~I\~niguez$^{5,6,11}$, G.~G.~Kenning$^{7}$,
A.~Maiorano$^{12,6}$, E.~Marinari$^{13}$,
V.~Martin-Mayor$^{8,6}$, J.~Moreno-Gordo$^{6,5,10}$, A.~Mu\~noz-Sudupe$^{8,6}$, D.~Navarro$^{14}$,
R.~L.~Orbach$^{15}$, G.~Parisi$^{13}$, S.~Perez-Gaviro$^{5,6}$, F. Ricci-Tersenghi$^{13}$,
J.~J.~Ruiz-Lorenzo$^{16,10,6}$, S.~F.~Schifano$^{17}$, D. L. Schlagel$^{18}$, B.~Seoane$^{6,20}$,
A.~Tarancon$^{5,6}$, D.~Yllanes$^{19,6}$.}

\affiliation{$^1$ Institute of Nanotechnology, Soft and Living Matter Laboratory, Consiglio Nazionale delle Ricerche (CNR-NANOTEC), Piazzale Aldo Moro 5, I-00185 Rome, Italy} 
\affiliation{$^2$ School of Physics, MOE Key Laboratory for Nonequilibrium Synthesis and Modulation of Condensed Matter, Xi’an Jiaotong University, Xi’an 710049, China}
\affiliation{$^3$ Eawag,  Überlandstrasse 133, CH-8600 Dübendorf, Switzerland}
\affiliation{$^4$ Dipartimento di Fisica e Scienze della
  Terra, Universit\`a di Ferrara e INFN, Sezione di Ferrara, I-44122
  Ferrara, Italy}
\affiliation{$^5$ Departamento de F\'\i{}sica Te\'orica,
  Universidad de Zaragoza, 50009 Zaragoza,
  Spain}
\affiliation{$^6$ Instituto de Biocomputaci\'on y F\'{\i}sica de
  Sistemas Complejos (BIFI), 50018 Zaragoza, Spain}
\affiliation{$^{7}$Department of Physics and Astronomy, Indiana University of Pennsylvania, Indiana, Pennsylvania 15705, USA}
\affiliation{$^8$ Departamento de   F\'\i{}sica Te\'orica, Universidad Complutense, 28040 Madrid, Spain}
\affiliation{$^9$ Departamento de
  Ingenier\'{\i}a El\'ectrica, Electr\'onica y Autom\'atica, U. de
  Extremadura, 10003, C\'aceres, Spain}
\affiliation{$^{10}$ Instituto de
  Computaci\'on Cient\'{\i}fica Avanzada (ICCAEx), Universidad de
  Extremadura, 06006 Badajoz, Spain}
\affiliation{$^{11}$ Fundaci\'on ARAID, Diputaci\'on General de
  Arag\'on, 50018 Zaragoza, Spain}
\affiliation{$^{12}$ Dipartimento di Biotecnologie, Chimica e
  Farmacia, Università degli studi di Siena, 53100, Siena,
  Italy and INFN, Sezione di Roma 1, 00185 Rome,
  Italy}
  \affiliation{$^{13}$ Dipartimento di Fisica, Sapienza
  Universit\`a di Roma, and CNR-Nanotec, Rome unit and
  INFN, Sezione di Roma 1, 00185 Rome, Italy}
\affiliation{$^{14}$ Departamento de Ingenier\'{\i}a,
  Electr\'onica y Comunicaciones and I3A, U. de Zaragoza, 50018
  Zaragoza, Spain}
  \affiliation{$^{15}$ Texas Materials Institute, The University of Texas at Austin, Austin,
  Texas 78712, USA}
 \affiliation{$^{16}$ Departamento de F\'{\i}sica,
  Universidad de Extremadura, 06006 Badajoz,
  Spain}

\affiliation{$^{17}$ Dipartimento di Scienze dell'Ambiente e della Prevenzione Università di Ferrara e INFN Sezione di Ferrara, I-44122 Ferrara, Italy}
\affiliation{$^{18}$ Division of Materials
  Science and Engineering, Ames Laboratory, Ames, Iowa 50011, USA}
\affiliation{$^{19}$ Chan Zuckerberg Biohub, San Francisco, CA 94158, USA}
\affiliation{$^{20}$ Université Paris-Saclay, CNRS, INRIA Tau team, LISN, 91190 Gif-sur-Yvette, France}
\affiliation{$^*$ These authors contributed equally to this work.}

\date{\today}

\begin{abstract}
\centerline{}
\centerline {\bf ABSTRACT}
\centerline{}
\noindent  
The extended principle of superposition has been a touchstone of spin glass dynamics for almost thirty years.  The Uppsala group has demonstrated its validity for the metallic spin glass, CuMn, for magnetic fields $H$ up to 10 Oe at the reduced temperature $T_\mathrm{r}=T/\Tg = 0.95$, where $\Tg$ is the spin glass condensation temperature.  For $H > 10$ Oe, they observe a departure from linear response which they ascribe to the development of non-linear dynamics. The thrust of this paper is to develop a microscopic origin for this behavior by focusing on the time development of the spin glass correlation length, $\xi(t,\tw;H)$.  Here, $t$ is the time after $H$ changes, and $\tw$ is the time from the quench for $T>\Tg$ to the working temperature $T$ until $H$ changes.  We connect the growth of $\xi(t,\tw;H)$ to the barrier heights $\Delta(\tw)$ that set the dynamics.  The effect of $H$ on the magnitude of $\Delta(\tw)$ is responsible for affecting differently the two dynamical protocols associated with turning $H$ off (TRM, or thermoremanent magnetization) or on (ZFC, or zero field-cooled magnetization). 
This difference is a consequence of non-linearity based on the effect of $H$ on $\Delta(\tw)$.
Superposition is preserved if $\Delta(\tw)$ is linear in the Hamming distance $\mathrm{Hd}$ (proportional to the
difference between the self-overlap $q_\mathrm{EA}$ and the overlap 
$q(\Delta(\tw))$). However, superposition is
violated if $\Delta(\tw)$ increases faster than linear in $\mathrm{Hd}$. We have previously shown, through
experiment and simulation, that the barriers $\Delta(\tw)$ do increase more rapidly than linearly with $\mathrm{Hd}$
through the observation that the growth of $\xi(t,\tw ;H)$ slows down as $\xi(t,\tw ;H)$ increases.
In this paper, we display the difference between the zero-field cooled $\xi_{\text {ZFC}}(t,\tw;H)$ and the thermoremanent magnetization $\xi_{\text {TRM}}(t,\tw;H)$ correlation lengths as $H$ increases, both experimentally and through numerical simulations, corresponding to the violation of the extended principle of superposition in line with the finding of the Uppsala Group.
	
\end{abstract}


\maketitle
\section{Introduction}
\label{sec:intro}
The dynamics of spin glasses have had impact on a variety of other physical systems~\citep{parisi:23}, and now even extends into the social sciences~\citep{cui:20}.  

One of the precepts of the dynamics is the so-called \textit{extended principle of superposition}, first introduced in the papers of the Uppsala group~\citep{nordblad:86,lundgren:86,nordblad:87,djurberg:95}.  In brief, it sets
\begin{equation}\label{eq:superposition_M}
M_{\text {TRM}}(t, \tw) + M_{\text {ZFC}}(t, \tw) = M_{\text {FC}}(0,\tw+t) \,.
\end{equation}
Here, $\tw$ is the time after the spin glass has been quenched to the working temperature $T$ from above the condensation temperature $T_g$ until the magnetic field $H$ is changed, and $t$ is the observation time after the change in $H$.  $M_{\text {TRM}}(\tw,t)$ denotes the thermoremanent magnetization: the spin glass magnetization for a temperature  quench to the working temperature $T<T_g$ in the presence of a magnetic field $H$, at the time $t$ after $H$ is cut to zero after the ``waiting time" $\tw$.  $M_{\text {ZFC}}(\tw,t)$ denotes the zero-field cooled magnetization: the spin glass magnetization for a temperature  quench in the absence of a magnetic field $H$, at time t after $H$ is applied after $\tw$.  Finally, $M_{\text {FC}}(0,\tw+t)$ is the field cooled magnetization, recorded continuously during the temperature quench to $T$ and during the sum of the times $\tw+t$.

While Eq.~\eqref{eq:superposition_M} seems satisfying, the Uppsala group noted that~\citep{nordblad:87}: ``...$M_{\text {ZFC}}(h)$ exhibits a linear field dependence...[and]...a nonlinear field dependence of $M_{\text {TRM}}(h)$...."  This would lead to a violation of the principle of superposition as the magnetic field ($h\equiv H$) increases.  Indeed, they observed violation for both a metallic 5 at.\% CuMn sample, at $H > 10$ Oe~\citep{nordblad:87}, and for an amorphous metallic spin glass $(\mathrm{Fe}_{0.15}\mathrm{Ni}_{0.85})_{75}\mathrm{P}_{16}\mathrm{B}_6\mathrm{Al}_3$ at $H>1$ Oe~\citep{djurberg:95}.

In addition, deviations from linear response was observed in the Ising spin glass
$\mathrm{Fe}_{0.5}\mathrm{Mn}_{0.5} \mathrm{TiO}_3$
by Ito et al.~\cite{ito:86} and in $\mathrm{Cu}_{13.5} \mathrm{Mn}_{86.5}$ by Hudi et al.~\cite{hudi:16}. In both these works, using lower and lower magnetic fields, the linear response regime
was entered, yielding identical relaxation functions for ZFC and TRM protocols, and hence,
satisfaction of the superposition principle, Eq. (\ref{eq:superposition_M}).

In the work described in this paper, we shall present the microscopic origin of this difference between $M_{\text {ZFC}}(\tw,t)$ and $M_{\text {TRM}}(\tw,t)$ as $H$ increases, leading to the breakdown of superposition, through experiment and numerical simulations. Our analysis will utilize the spin glass correlation length $\xi(t,\tw;H)$ that grows from nucleation after the temperature quench to $T<T_g$.

The growth rate of $\xi(t,\tw;H)$ at the time $t$ is set by the free energy barriers $\Delta(\tw)$ created by its growth during the waiting time $\tw$. The \emph{separation} between states will be denoted by the Hamming Distance ($\mathrm{Hd}$), and is defined below in Appendix ~\ref{appendix:phenomenological_HD}, Eq.~\eqref{eq:Hdistance_def}.  We shall show that, if the relationship is linear, Eq.~\eqref{eq:superposition_M} holds.  If $\Delta(t,\tw)$ increases more rapidly with $\mathrm{Hd}$ than linear, Eq.~\eqref{eq:superposition_M} is violated.  The decay of the TRM will then be slower than the rise of the ZFC, and the departure from Eq.~\eqref{eq:superposition_M} will increase with increasing magnetic field change and increasing $\tw$. That $\Delta(t,\tw)$ increases with increasing $\xi(t,\tw;H)$ has already been displayed 
by us~\citep{zhai:19}, where we showed that the growth of $\xi(t,\tw;H)$ slows down as $\xi(t,\tw;H)$ grows.

Our analysis will show that the experimental method for extracting $\xi$ is more dependable in the ZFC protocol than in the TRM setting. Indeed, we shall show that $\xi$, as obtained for both protocols, behaves differently  with time in a field.
Nevertheless, the two protocols become equivalent in the limit of a vanishing magnetic field. The same conclusion is reached from the microscopic computation of $\xi$ that is discussed in Sec.~\ref{sec:xi_num}, below.

The remaining part of this work is organized as follows.  The next Section will introduce a phenomenological approach based on the Hamming distance for outlining the effect of the magnetic field on spin-glass dynamics.  Section \ref{sec:difference_xi_in_exp} will exhibit experimental results for $\xi_{\text {ZFC}}(t,\tw;H)$ and $\xi_{\text {TRM}}(t,\tw;H)$ for a CuMn 8 at.\% single crystal sample.  Section ~\ref{sec:dependence_Delta_vs_Hd_exp} will present a quantitative relationship for the dependence of the free energy barrier $\Delta(\tw)$ on $\mathrm{Hd}$ from these experiments.

Section ~\ref{sec:xi_num} will present the results of simulations on a massive special purpose supercomputer, Janus II, for both $\xi_{\text {ZFC}}(t,\tw;H)$ and $\xi_{\text {TRM}}(t,\tw;H)$.  Section \ref{subsec:details_num} will give details of the simulations; Section ~\ref{subsec:violation_superpostion} will give the origin, and display the failure, of the extended superposition principle at finite $H$.

Section~\ref{sec:diff_TRM_ZFC_num} displays the difference between the TRM and ZFC protocol from a microscopic point of view. In Section \ref{subsec:num_approachZFC_TRM_xi_micro}, we evaluate the microscopic value for $\xi_{\text {micro}}(t,\tw;H)$ through the replicon progator, comparing ZFC with TRM values; and in Section~\ref{subsec:exp_approachZFC_TRM_teff} we unveil the difference between the two protocols through the lens of the magnetic response.  In Section \ref{subsec:ximicro_vs_xieff} we define an \emph{effective} correlation length, proving that it measures quantities equivalent to the microscopical correlation length. Section \ref{sec:conclusion} summarizes the conclusions of this paper, and points to future investigations based on these results.  The paper concludes with eight appendixes.

\section{A phenomenological approach based on the Hamming distance}
\label{sec:phenomenological_HD}

The beautiful solution of the \emph{mean-field} Sherrington-Kirkpatrick
model~\cite{mezard:84} is a landmark in the physics of disordered systems and
gives rise to a theoretical picture that is known by several names, such as
the \emph{Replica Symmetry Breaking} or the \emph{hierarchical} models of spin
glasses. Unfortunately, connecting this solution to experimental work is not
straightforward, because of two major difficulties.  First, strictly speaking,
the mean-field solution applies only to space dimension higher than six (alas,
experiments are carried out in the three-dimensional world in which we live). How this hierarchical picture needs to be modified in three dimensions is a
much-debated problem (see e.g., Ref.~\cite{martin-mayor:22} for an updated
account). The second, and perhaps more serious problem, is related to the fact that this theory describes systems in thermal equilibrium. Now, below the
glass temperature, the correlation length $\xi$ of a system in thermal
equilibrium is as large as the system's size. Unfortunately, because of the
extreme slowness of the time growth of $\xi$, the experimental situation is
the opposite: the sample size is typically much larger than $\xi$. Clearly,
some additional input is needed to connect the hierarchical picture of the
spin-glass phase with real experiments.

One such connecting approach is based on generalized fluctuation-dissipation
relations~\cite{cugliandolo:93,franz:95,marinari:98f,franz:98,franz:99,marinari:00b,cruz:03,janus:17},
which have been also investigated experimentally for atomic spin
glasses~\cite{herisson:02,herisson:04}. Unfortunately, these relations focus
on the \emph{linear} response to the magnetic field, while non-linear relations
will be crucial to us.

An alternative approach was worked out in Ref.~\cite{joh:96}, which explores the dynamics in an ultrametric tree of states. A crucial quantity in this approach is the Hamming distance ($\mathrm{Hd}$) between the state of the system after the initial preparation at time $\tw$, and the state at the measuring time $t+\tw$.  Yet, we still do not know how this Hamming distance should be defined microscopically. We do have a surrogate that can be obtained from a correlation function (this correlation function can be computed, see
Appendix.~\ref{subsec:Hd_connection_to_Ez_num}, and experimentally
measured~\cite{herisson:02}). Unfortunately, the surrogate is not a fully
adequate substitute for the Hamming distance of Joh \emph{et
  al.}~\cite{joh:96}. Nevertheless, the dynamics in the hierarchical tree does
provide useful intuition. This is why we briefly recall here its main results
and assumptions.  The interested reader will find a more complete account in
Appendix~\ref{appendix:phenomenological_HD}. In fact, we shall take a further
step because, at variance with Ref.~\cite{joh:96}, we shall accept the
possibility that barrier heights increase faster than linearly with Hd (we
shall work out the consequences of this possibility as well).

There are many experimental protocols for exploring spin-glass dynamics.
As we explained in the Introduction, those that involve the time change of the
magnetization are the zero field cooled magnetization (ZFC) and the
thermoremanent magnetization (TRM) protocols, generating
$M_\mathrm{ZFC}(t,\tw;H)$ and $M_\mathrm{TRM}(t,\tw;H)$, respectively. The
basic concept in the analysis will be the maximum free-energy barrier
$\varDelta_{\text{max}}$ between the involved states (see
Appendix~\ref{appendix:phenomenological_HD}). Take, for instance, the TRM
protocol. When the magnetic field $H$ is cut to zero, the system remembers its
correlations achieved after aging for the time $\tw$.  This generates an
inflection point in the time decay of $M_\mathrm{TRM}(t,\tw;H)$ at
$t\approx \tw$.  This is exhibited as a peak in the relaxation
function~\cite{granberg:88},
\begin{equation}\label{eq:relaxationS_def}
S(t,\tw;H)_\mathrm{(ZFC/TRM)} = (\pm) {\frac {	\mathrm{d} \, M_\mathrm{TRM}(t,\tw;H)}{\mathrm{d} \,\log\,t}} \; ,
\end{equation}
where the $-$ sign pertains to TRM experiments and $+$ sign to ZFC experiments.
The log of the time at which $S(t,\tw;H)$ peaks, $\log \teff_H$,\footnote{Here, and all over the text, $\log$ is the natural logarithm; otherwise,  the logarithmic basis is explicitly indicated, \textit{i.e. $\log_2$}.} is thus a
measure of $\varDelta_\mathrm{max}$. The activation time is set approximately
by the maximum barrier height reached in the waiting time $\tw$:
\begin{equation}\label{eq:teff_arrehenius_law}
\teff_H=\tau_0\,\mathrm{e}^{\varDelta_\mathrm{max}(t=0,\tw;H)/k_B T} \; ,
\end{equation}
where $\tau_0$ is an exchange time of the order of $\hbar/k_B T_\text{g}$, and
$\varDelta_\mathrm{max}(t=0,\tw;H)$ is the highest barrier created by the growth
of $\xi(t=0,\tw;H)$ in the time $\tw$.

As Bouchaud has shown~\cite{bouchaud:92}, when a magnetic field is present the barrier heights $\varDelta$ are reduced by a Zeeman energy $E_\mathrm{Z}$:
\begin{equation}\label{eq:Delta_reduction_Ez}
\varDelta(t,\tw;H) = \varDelta(t,\tw;0) + E_\mathrm{Z}~~ .
\end{equation}
For small magnetic field, $E_\mathrm{Z}$ behaves as ~\cite{joh:99,guchhait:14,bouchaud:92,vincent:95,janus:17b}:
\begin{equation}\label{eq:zeeman_energy_def}
E_\mathrm{Z}= - M_\mathrm{FC} H \equiv - \chi_\mathrm{FC} N_\mathrm{c}H^2~~.
\end{equation}
where, $\chi_\mathrm{FC}$ is the field-cooled magnetic susceptibility {\it per spin} when the spin glass is cooled in a magnetic field to the measurement temperature $T$; $N_\mathrm{c}$ is the number of correlated spins [spanned by the spin glass correlation length, $\xi_\mathrm{Zeeman}(t,\tw;H)$]
\begin{equation}\label{eq:Nc_definition}
N_\mathrm{c} \approx \xi_\mathrm{Zeeman}^{3-\theta/2}~~,
\end{equation}
$\theta$ is the replicon exponent~\cite{janus:17b} (see also Appendix \ref{appendix:scaling}), and $H$ is the applied  magnetic field.\footnote{
Another view of $E_\mathrm{Z}$ (Bert \emph{et al.}~\cite{bert:04}) relies on
fluctuations in the magnetization of all of the spins.  They used
$E_\mathrm{Z}$ linear in $H$, replacing $N_\mathrm{c}$ with $\sqrt
N_\mathrm{c}$, and using the free spin value in place of $\chi_\mathrm{FC}$. A
very recent investigation of the magnetic field's effect on spin-glass dynamics,
Paga \emph{et al.}~\cite{zhai-janus:21}, shows that their fit to experiments
can also be ascribed to non-linear effects introduced by their use of rather
large values of the magnetic field. We therefore shall use Eq.~\eqref{eq:zeeman_energy_def} in our subsequent analysis.}

\section{\boldmath $\xi_\mathbf{TRM}(t, \tw;H)$ and $\xi_\mathbf{ZFC}(t, \tw;H)$ from experiment}
\label{sec:difference_xi_in_exp}

In this section we will provide  details of the experiments and some theoretical background. After that, we will describe the computation of  $\xi_\mathbf{TRM}(t,\tw;H)$ and $\xi_\mathbf{ZFC}(t,\tw;H)$ from experiments.

\subsection{Details of the experiments}

The TRM and ZFC experiments were performed on 8 at.\% CuMn samples cut from the same single-crystal boule grown at Ames Laboratory, and characterized in Ref.~\cite{zhai:19}.  The TRM experiments were performed both at the Indiana University of Pennsylvania on a home-built SQUID magnetometer, capable of sensitivity roughly an order of magnitude greater than commercial devices, and at The University of Texas at Austin on a Quantum Design commercial SQUID magnetometer.  The ZFC experiments were performed at The University of Texas at Austin on the same equipment as the TRM.  

\subsection{Some analytical background}

Both ZFC and TRM protocols use the time at which $S(t,\tw;H)$ peaks to be the measure of $\teff_H$,  see  Eq.~\eqref{eq:relaxationS_def} for $S(t,\tw;H)$ definition. The measurements were all made at $37.5$~K, or a reduced
temperature ($T_\mathrm{g}=41.5$ K) of $T_\mathrm{r}=0.9$. Two waiting times were set at $\tw=
2500$~s and $5000$~s, testing the growth 
law~\cite{marinari:96,kisker:96,sibani:94}
\begin{equation}\label{eq:xi_growth_law}
\xi(\tw)=c_1\,\bigg({\frac {\tw}{\tau_0}}\bigg)^{c_2(T/T_\mathrm{g})}\,,
\end{equation}
where $c_1$ is a constant of order unity and $c_2\approx 0.104$.
The time $\teff_H$ at which $S(t,\tw;H)$ peaks is
indicative of the largest barrier $\varDelta_\mathrm{max}(\tw;H)$ surmounted in
the time $\tw$~\cite{nordblad:86}. 
In the $H\rightarrow 0$ limit, $S(t,\tw;H)$ peaks close to $\tw$.  The
{\it shift} of the peak of the relaxation function $S(t,\tw;H)$ from $\tw$ to $\teff_H $ as $H$ increases from zero is a direct measure of the reduction of
$\varDelta_\mathrm{max}$ with increasing $H$ [see
Eq.~\eqref{eq:Delta_reduction_Ez}]. Thus, combining Eq.~\eqref{eq:teff_arrehenius_law} with Eq.~\eqref{eq:Delta_reduction_Ez}:
\begin{equation}\label{eq:Delta_Ez_reduction_vs_teff}
\varDelta_\mathrm{max} - N_\mathrm{c} \chi H^2=k_\mathrm{B} T[\log\,\teff_H-\log\,\tau_0]\,, 
\end{equation}
we can estimate the number of correlated spins, [i.e $\xi_\mathrm{Zeeman}(\tw)$], through the decay of the effective time $\teff_H$ 
\begin{equation}\label{eq:xi_Zeeman_through_teff}
\log(\teff_{H}) \propto \xi_\mathrm{Zeeman}^{3-\theta/2} H^2 \, .
\end{equation}
The above expression exploits the formalism  introduced by Joh 
\emph{et al.}~\cite{joh:99}, which had a further development in Refs.~\cite{zhai-janus:20a,zhai-janus:21} considering scaling theory and non-linear effects as:
\begin{equation}\label{eq:scaling_law_decay_teff}
\begin{split}
\log \bigg[{\frac {\teff_H}{t_{H\rightarrow 0^+}^{\text {eff}}}}\bigg] &={\frac {\hat{S}}{2T}}\xi(\tw)^{D-(\theta/2)}H^2\\
& +\xi(\tw)^{-\theta/2}{\mathcal G}\big(T,\xi(\tw)^{D-(\theta/2)}H^2\big) \, .
\end{split}
\end{equation}
Here, $\xi(\tw)$ is the spin-glass correlation length,  $\hat{S}$ is a constant from the Fluctuation-Dissipation Theorem (FDT), $D=3$ is the spatial dimension, and $\mathcal{G}(x)$ is a scaling function behaving as $\mathcal{G}(x) \sim x^2$ for small $x$. The replicon exponent $\theta$ is a function of $\tilde{x}= \ell_\mathrm{J}(T)/\xi(\tw)$, where $\ell_J(T)$ is the Josephson length~\cite{janus:18,zhai:19}. For notational simplicity, we have omitted this functional dependence.

As the reader can notice, Eq.~\eqref{eq:xi_Zeeman_through_teff} neglects the $\mathcal{O}(H^4)$ terms in Eq.~\eqref{eq:scaling_law_decay_teff}, and we have explicitly written $\xi_\mathrm{Zeeman}$ in Eq.~\eqref{eq:xi_Zeeman_through_teff} since we are referring to experiments. The general expression, Eq.~\eqref{eq:scaling_law_decay_teff}, will be treated in  numerical sections.

Refs.\cite{zhai-janus:20a,zhai-janus:21} demonstrated how to connect experiments to simulations, and vice versa, despite the range of time and length scales, and the external magnetic fields.
Further proof of these equivalences is the first numerical evidence of the exotic phenomena of rejuvenation and memory \cite{janus:23}.

\subsection{Checking \boldmath $\xi_\mathbf{TRM}(t,\tw;H)< \xi_\mathbf{ZFC}(t,\tw;H)$}

In this manner, the TRM and ZFC protocols generate $\xi_\mathrm{TRM}(t,\tw;H)$ and $\xi_\mathrm{ZFC}(t,\tw;H)$, respectively.  The hypothesized difference, $\xi_\mathrm{TRM}(t,\tw;H) < \xi_\mathrm{ZFC}(t,\tw;H)$, can then be tested.

We expect that difference, if any, to be a result of an upward curvature of
$\varDelta$ as a function of $\mathrm{Hd}$, as outlined in the previous Section
and in Appendix~\ref{appendix:phenomenological_HD}.
Fig.~\ref{fig:log_teff_375exp} exhibits experimental values of
$\log\,\teff_H$ vs $H^2$ for the 8 at.\% $\mathrm{CuMn}$ sample, and fits to the data for the ZFC and TRM
protocols for waiting times $\tw=2500$~s and $\tw=5000$~s at $T=37.5$ K.
Because $T=37.5$ K is so close to $T_\mathrm{g}$, non-linear terms are evident
in the data. As a consequence, the fits employ higher-order terms in $H$ as well as quadratic.

Appendix~\ref{appendix:second-single-crystal} presents data taken from
another sample cut from a 6 at.\% CuMn single crystal boule with
$\Tg = 31.5$ K at a measurement temperature of $\Tm = 26$ K.  The
growth of $\xi(t,\tw; H)$ was slower for both ZFC and TRM protocols,
leading to smaller values of the correlation lengths and therefore
smaller differences between $\xi_{\text {TRM}}(\tw)$ and $\xi_{\text
  {ZFC}}(\tw)$, as compared to those exhibited here in the main text.
Nevertheless, at the largest waiting time, the difference lies well
outside the sum of the error bars.

\begin{figure}[h]
	\centering
	\includegraphics[width = 1\columnwidth]{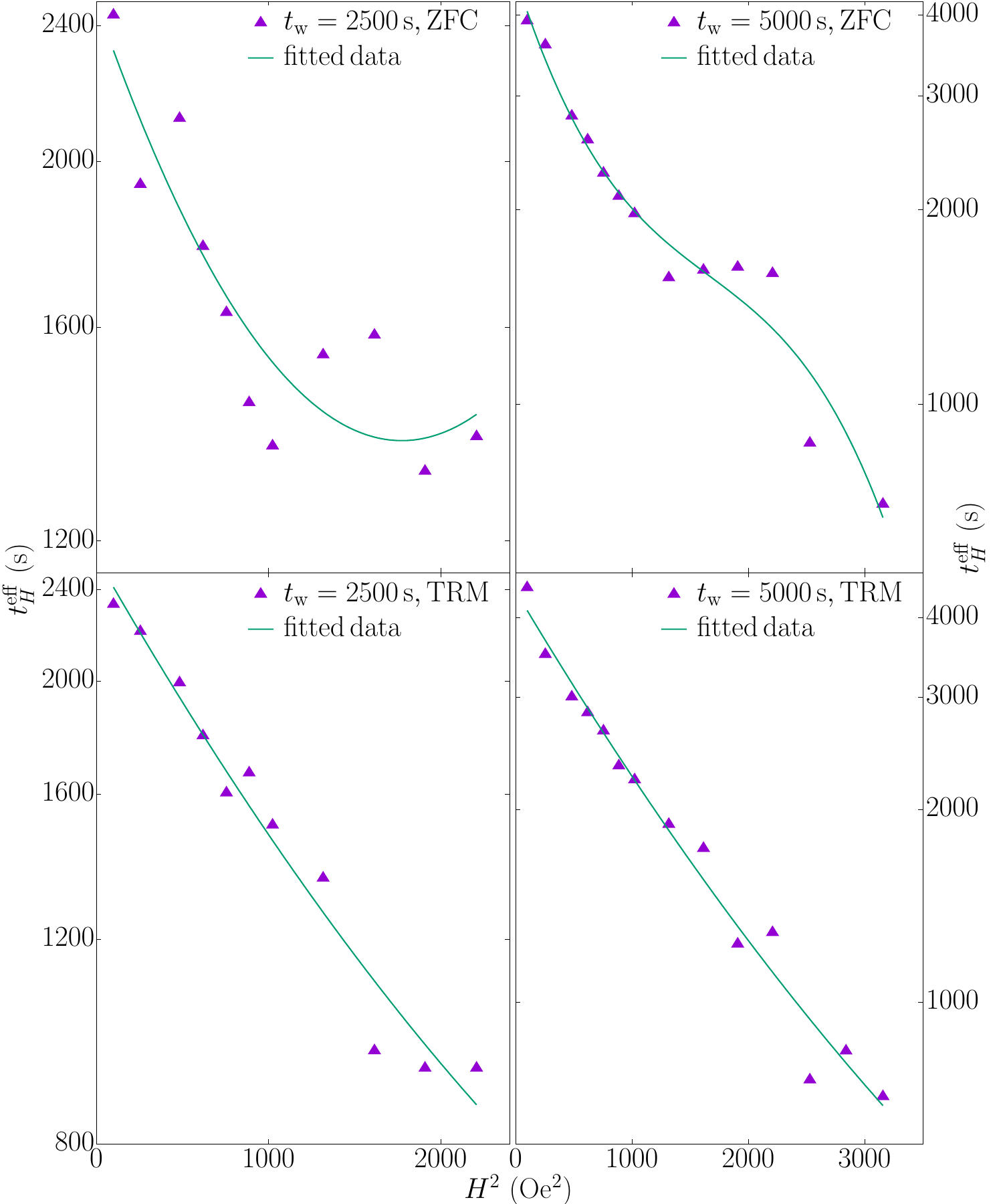}
	\caption{A plot of data and fit for the effective
waiting time $\teff_H$ (logarithmic scale) vs $H^2$ for ZFC and TRM measurements on a CuMn 8 at.\% single crystal at $\tw=2500, 5000$~s and $T=37.5$~K.  The polynomial fitting
parameters for each of the values of $H$, up to and including $H^4$, are given
in Tables~\ref{tab:zfc375t0}---\ref{tab:trm375} (see Appendix~\ref{appendix:only_tables}).  The value of the correlation length  is extracted from the $H^2$ fitting terms.}
	\label{fig:log_teff_375exp}
\end{figure}

Fitting to the coefficients of the $H^2$ terms for the 8 at.\% CuMn sample described above (see the respective tables in Appendix~\ref{appendix:only_tables}
for specific values), we are able to extract
$\xi_\mathrm{ZFC}(\tw)$ and $\xi_\mathrm{TRM}(\tw)$.\\
We find:
\begin{align}
\nonumber \xi_\mathrm{ZFC}(\tw=2500\text{ s})&=220(20), \\
\nonumber \xi_\mathrm{TRM}(\tw=2500\text{ s})&=210(16), \\
\xi_\mathrm{ZFC}(\tw=5000\text{ s})&=270(20),  \label{eq:values_xi_exp} \\
\nonumber \xi_\mathrm{TRM}(\tw=5000\text{ s})&=220(30),
\end{align}
all in units of the lattice constant $a_0$.  As hypothesized, the magnitude of
$\xi_\mathrm{ZFC}(\tw)$ exceeds $\xi_\mathrm{TRM}(\tw)$. It must be
noted, however, that the difference lies well within the error bars for $\tw=2500$~s,
while the difference is just inside the sum of the error bars for $\tw=5000$~s.

Our attempts at larger values of $\tw$ have not been successful, as the
$S(t,\tw;H)$ curves broaden so much that it proved too difficult to extract
reproducible values for $\teff_H$.  Smaller values of $\tw$ were not
attempted as the difference between the ZFC and TRM correlation lengths would be
smaller than for $\tw=2500$~s and the error bars would obviate any reliable
conclusions.

The ratio for the respective values of $\xi(t,\tw;H)$ as a function of waiting time, $\tw$, is instructive. One finds
\begin{align*}
\xi_\mathrm{ZFC}(\tw\!=\!5000\text{ s})/\xi_\mathrm{ZFC}(\tw\!=\!2500\text{ s})&=1.26 \\
\xi_\mathrm{TRM}(\tw\!=\!5000\text{ s})/\xi_\mathrm{TRM}(\tw\!=\!2500\text{ s})&=1.06 \; . 
\end{align*}
This confirms that $\xi_\mathrm{TRM}(t,\tw;H)$ grows more slowly
than $\xi_\mathrm{ZFC}(t,\tw;H)$ with $\tw$, indicating that the barrier
heights  encountered by $\xi(t,\tw,H)$ are larger for TRM growth.  This is additional evidence that the
dependence of $\varDelta(\tw)$ on $\mathrm{Hd}$ curves upward.

All the measurements reported above were made with the magnetic field $H$
calibrated using a standard Pd sample provided by Quantum Design. The residual
magnetic field was measured before each series of measurements, and the cooling
profile kept the same for both ZFC and TRM measurements. The self-consistency
of each data set, when combined with calibration of $H$ as described above,
gives some confidence in the differences between $\xi_\mathrm{TRM}(t,\tw;H)$
and $\xi_\mathrm{ZFC}(t,\tw;H)$.

\section{\boldmath Explicit dependence of $\varDelta(\tw)$ on Hamming distance ($\mathrm{Hd}$)}
\label{sec:dependence_Delta_vs_Hd_exp}

We have organized this section as follow.  First, we introduce some experimental historical background
on the relation between barriers and the Hamming distance, and then we shall describe our findings from our experiments.

\subsection{Some background}

The paper ``Dynamics in spin glasses'' by Lederman \emph{et al.}~\cite{lederman:91}, as a consequence of detailed experiments on AgMn (2.6 at.\%), developed a quantitative relationship between the change in $\varDelta(\tw)$ as a function of the change in $\mathrm{Hd}$ \footnote{The only reference in the literature that contains a quantitative relationship between the change in the barrier height encountered in  a waiting time $\tw$ and the Hamming distance is for AgMn. As a consequence, we have used their relationship in our analysis for CuMn. 
The point of the analysis is to demonstrate that for a doubling of the waiting time $\tw$, there is hardly a change in $\mathrm{Hd}$.  This, of course, is representative of the remarkably slow dynamics encountered in spin glasses.  If there is a slight difference in the parameters extracted from experiments for AgMn as compared to CuMn, the actual values of $\mathrm{Hd}$ would change, but not their order of magnitude.}.  Writing out their Eq. (13), [see our Eq.~\eqref{eq:variation_Delta_vs_Hd}]
\begin{equation}\label{eq:Lederman_eq13}
\begin{split}
\varDelta(\mathrm{Hd})-\beta(T)/\alpha(T)\\
=[\varDelta(\mathrm{Hd}_0)-\beta(T)&/\alpha(T)]e^{[\alpha(T)(\mathrm{Hd}-\mathrm{Hd}_0)]},
\end{split}
\end{equation}
where $\varDelta(\mathrm{Hd})-\varDelta(\mathrm{Hd}_0)$ is the change in barrier heights when $\mathrm{Hd}$ increases from $\mathrm{Hd}_0$ to $\mathrm{Hd}$.  The coefficient in the exponent, $\alpha(T)$, was estimated from experiment to be approximately 38.1 at a reduced temperature of $T_\text{r}=0.865$.  Further, $\alpha(T)$ and $\beta(T)$ are defined by:
\begin{equation}\label{eq:alpha_beta_expression}
\begin{split}
\alpha(T) &= -2a(T)/(\delta q_\mathrm{EA}/\delta T), \\ \beta(T)&=-2b(T)/(\delta q_\mathrm{EA}/\delta T),
\end{split}
\end{equation}
where $q_\mathrm{EA}$ is the Edwards-Anderson  self-overlap, and $a(T)$ and $b(T)$ are defined by an experimental fit,
\begin{equation}\label{eq:Delta_over_deltaT}
(\delta\varDelta/\delta T)|_{T}=a(T)\varDelta+b(T) \, .
\end{equation}
Figs. 11 and 12 of Ref.~\cite{lederman:91}  display $a(T)$ and $b(T)$, respectively, for four
representative values of $T$.  For our purposes, we are only interested in
the ratio $\beta/\alpha$, which, from Eq.~\eqref{eq:alpha_beta_expression},
is independent of $\delta q_\mathrm{EA}/\delta T$. Our working
temperature is $T = 37.5$, or a reduced temperature $T_\mathrm{r}=0.90$.  From Fig. 10
of Lederman \emph{et al.}~\cite{lederman:91}, this leads to $a\approx 29.02$
and $b\approx 684.03$, generating the ratio,
\begin{equation}\label{eq:ratio_ab}
\beta/\alpha=b/a\approx 23.57.
\end{equation} 

\subsection{Correlation lengths and  Hamming distances from our experiments}

On the assumption that the ratios for AgMn are relevant to CuMn, we can then address our data.  At $T = 37.5$ K, we fit the time at which $S(t,\tw;H)$ peaks, $\teff_H$ by,
\begin{equation}\label{eq:fitting_for_teff}
\log(\teff_H)=a_0+a_2H^2+a_4H^4+a_6H^6+\mathcal{O}(H^8)\,.
\end{equation}
Eq.~\eqref{eq:fitting_for_teff} can be converted to an energy scale by rewriting as
\begin{equation}\label{eq:teff_explicit_coefficients_dep}
\begin{split}
k_\mathrm{B} T\log(\teff_H/\tau_0)=k_\mathrm{B}T[a_0-\log(\tau_0)]~~~~~~~~~~~~~~~~~\\
+k_\mathrm{B} T[a_2H^2+a_4H^4+a_6H^6+\mathcal{O}(H^8)]\,.\\
\end{split}
\end{equation}
Dividing Eq.~\eqref{eq:teff_explicit_coefficients_dep} by $k_\mathrm{B} T_\mathrm{g}$
gives the energy scale in units of $k_\mathrm{B} T_\mathrm{g}$.  We define
$\varDelta_0(\tw)\equiv E_0 = (T/T_\mathrm{g})[a_0-\log(\tau_0)]$ as the height of
the last barrier encountered during a waiting time $\tw$ in the absence of a
magnetic field, and 
\begin{equation}
\label{eq:defEn}
E_n(\tw;H) = (T/T_\mathrm{g})a_n(\tw)H^n\,, 
\end{equation}
as the nth-order
change in the barrier height's free-energy scale caused by the presence
of the external magnetic field at the waiting time $\tw$.  The ZFC experiments
for $\tw=2500$~s yields $\varDelta_0^\mathrm{ZFC}(\tw=2500\text{ s})\equiv E_0=33.55848$ in
units of $k_\mathrm{B}T_\mathrm{g}$ (see Table~\ref{tab:zfc375t0} in Appendix~\ref{appendix:only_tables}).  The
value for the TRM experiments is $\varDelta_0^\mathrm{TRM}(\tw=2500\text{ s})\equiv
E_0=33.58718$ (see Table~\ref{tab:trm375t0}), which should be the same as
for the ZFC protocol, the slight difference being a result of  fits to the
data.  Likewise, for $\tw=5000$~s, Tables~\ref{tab:zfc375} and \ref{tab:trm375}
give $\varDelta_0^\mathrm{ZFC}(\tw=5000\text{ s})\equiv E_0=34.11658$ and
$\varDelta_0^\mathrm{TRM}(\tw=5000\text{ s})\equiv E_0=34.07752$, respectively.

From these values, and Eqs.~\eqref{eq:Lederman_eq13} and \eqref{eq:ratio_ab} with $\alpha(T_\mathrm{r}\!=\!0.90)=46.97$, we can arrive at $\delta \mathrm{Hd}\equiv \mathrm{Hd}-\mathrm{Hd}_0$.  We find
\begin{equation}\label{eq:Hd_delta_variation}
\begin{split}
\delta \mathrm{Hd}^{\text {TRM}}= 1.02 \times 10^{-3},\\
\delta \mathrm{Hd}^{\text {ZFC}}= 1.16 \times 10^{-3}.
\end{split}
\end{equation}
The small values of the difference in Hamming distances for a doubling of the
waiting times is an indication of the slow growth of the correlation lengths
with waiting times $\tw$.  The equilibrium value of the Hamming distance
[$q_{\alpha\beta}=0$ in Eq.~\eqref{eq:Hdistance_def}] is approximately 0.0575 (see below)
so, even for $\tw = 5000$~s, the change in Hamming distance is still tiny.  To
reach equilibrium would indeed require time scales of the order of the age of
the universe.

One can relate the  correlation length $\xi(\tw)$ directly to $\mathrm{Hd}$.  As
shown above, the Hamming distance increases by unity for each mutual spin flip,
that is, for each reduction in $q_{\alpha\beta}$ by two.  Thus, $\xi(\tw)$
increases by a lattice constant for each mutual spin flip.  The volume of real
space increases as $[\xi(\tw)]^{D-\theta/2}$, where $D$ is the spatial
dimension.  This must equal the total number of mutual spin flips, given by
$N\times \mathrm{Hd}(\tw)$. See Appendix \ref{appendix:scaling} for a derivation of this equation using renormalization group arguments.

Tables \ref{tab:zfc375t0}---\ref{tab:trm375}, see Appendix~\ref{appendix:only_tables}, give the
correlation lengths $\xi(\tw)$ for $\tw=2500$ and $\tw=5000$~s.  From
Eq.~\eqref{eq:Hd_delta_variation} the change in $\mathrm{Hd}$ is known for both
waiting times.  One can therefore take the ratio of $\xi(\tw)$ for the two
waiting times for each protocol, and establish an absolute value for
$\mathrm{Hd}(\tw)$ at each value of $\tw$.  Expressed numerically,
\begin{align}
\label{eq:estimation_Hd_through_xi_ratio}
\frac {\xi(\tw=5000\text{ s})^{[D-(\theta(t_\mathrm{w}\!=\!5000\text{ s})/2)]}}{\xi(\tw\!=\!2500\text{ s})^{[D-(\theta(t_\mathrm{w}\!=\!2500)/2)]}}&={\frac {\mathrm{Hd}_{5000}}{\mathrm{Hd}_{2500}}}\\
&={\frac {\mathrm{Hd}_{2500}+\delta \mathrm{Hd}}{\mathrm{Hd}_{2500}}} \; .\nonumber
\end{align}

In order to evaluate Eq.~\eqref{eq:estimation_Hd_through_xi_ratio}, it is necessary to know the respective values of $\theta$ at $T=37.5$ K.  They are
\begin{equation}
\begin{split}
\theta_\mathrm{ZFC}(T=37.5~{\text {K}}, t_\mathrm{w}=2500~{\text{s}})=0.354,\\
\theta_\mathrm{TRM}(T=37.5~{\text {K}}, t_\mathrm{w}=2500~{\text{s}})=0.356,\\
\theta_\mathrm{ZFC}(T=37.5~{\text {K}}, t_\mathrm{w}=5000~{\text{s}})=0.343,\\
\theta_\mathrm{TRM}(T=37.5~{\text {K}}, t_\mathrm{w}=5000~{\text{s}})=0.353.\\
\end{split}
\end{equation}
Using the values of $\xi(\tw)$ from Tables \ref{tab:zfc375t0}---\ref{tab:trm375}, and $\delta \mathrm{Hd}$ from Eq.~\eqref{eq:Hd_delta_variation}, one obtains
\begin{equation}\label{eq:Hd_values}
\begin{split}
\mathrm{Hd}_{\tw\!=\!2500}^{\text {TRM}}=5.64 \times 10^{-3},\\
\mathrm{Hd}_{\tw\!=\!5000}^{\text {TRM}}=6.66 \times 10^{-3},\\
\mathrm{Hd}_{\tw\!=\!2500}^{\text {ZFC}}=1.20 \times 10^{-3},\\
\mathrm{Hd}_{\tw=5000}^{\text {ZFC}}=2.36 \times 10^{-3} .
\end{split}
\end{equation}
The value of $q_{\mathrm{EA}}$ at $T_\text{r}=0.90$ is approximately
0.115, so that the full Hamming distance at this temperature, from
Eq.~\eqref{eq:Hdistance_def} with $q_{\alpha\beta}=0$, is 0.0575.  The
occupied phase space in our experiments from the results exhibited in
Eq.~\eqref{eq:Hd_values} therefore spans only about 12\% of the
available phase space.  The slow growth of $\xi(\tw)$ is evidence that
true equilibrium in the spin-glass condensed phase can never be
accomplished over laboratory time scales, except perhaps at temperatures
in the immediate vicinity of $T_\mathrm{g}$ where $q_\mathrm{EA}$ can
become small.

It is also interesting to note from Eq.~\eqref{eq:Hd_values} that
$\mathrm{Hd}$ for TRM experiments is larger than for ZFC experiments.
This is, of course, consistent with our picture of the shift of the
beginning of aging from $q_\mathrm{EA}$ for ZFC experiments to
$q_\mathrm{EA}-q(E_\mathrm{Z})$ or, equivalently from $\mathrm{Hd}=0$
to $\mathrm{Hd}(|E_\mathrm{Z}|=\varDelta)$ from Eq.~\eqref{eq:Hd_vs_Ez}
for TRM protocols as compared to ZFC protocols.

An important lesson from this analysis is that a true definition of $\xi(\tw)$
can be extracted only from a ZFC protocol.  A TRM protocol, assuming that
$\varDelta(\tw)$ increases with $\mathrm{Hd}$ faster than linearly, will generate
a value for $\xi(\tw)$ that is a function of the magnetic field.  In that
sense, though an average is usually taken, the only meaningful protocol for
extraction of $\xi(\tw)$ is ZFC.

In the next two sections (Secs.~\ref{sec:xi_num}-\ref{sec:diff_TRM_ZFC_num}), we will show results from numerical simulations in order to
\begin{enumerate}
\item Determine the microscopic features of a 3D spin glass in the presence of an external magnetic field;
\item Compute the difference between the magnetic response to the thermoremanent magnetization (TRM) and the zero-field-cooled (ZFC) protocols;
\item Probe the equivalence between the microscopic correlation length, $\xi_\mathrm{micro}(\tw)$, (calculated through the replicon propagator, see Appendix~\ref{appendix:xi_micro_def}) and the \emph{effective} correlation length, $\xi_\mathrm{Zeeman}$ (extracted through the lens of the magnetic response).
\end{enumerate}

\section{\boldmath $\xi_\mathrm{TRM}(t,\tw;H)$ and $\xi_\mathrm{ZFC}(t,\tw;H)$ from simulations}
\label{sec:xi_num}
This section is organized as follows.  In Sec.~\ref{subsec:details_num} we
present the details of the simulations carried out on the Janus~II
supercomputer. In  Sec.~\ref{subsec:violation_superpostion} we  shall assess the validity of the extended superposition principle, see
Eq.~\eqref{eq:superposition_M}. Finally, we
conclude this section with the extraction of an effective correlation length
for both the ZFC and the TRM protocols using the relationship relied upon
through experiment. We present a microscopic direct calculation of the correlation
length.

\subsection{Details of the simulations}
\label{subsec:details_num}

We have carried out massive simulations on the Janus II 
supercomputer~\cite{janus:14} studying the Ising-Edwards-Anderson (IEA) model on a cubic
lattice, with periodic boundary conditions and size $L=160~a_0$, where $a_0$ is
the average distance between magnetic moments.  
\begin{table}[!h]
\begin{centering}
\begin{ruledtabular}
\begin{tabular}{c c c c c c c}
&$\Tm$&$\tw$ & $\xi(\tw,H\!=\!0)$ & $t_{\text {max}}$ & $\theta(\tilde {x})$&$C_\mathrm{peak}(\tw)$\\ 
\hline\\[-5pt]
${\textbf{Run 1}}$&0.9&$2^{22}$&8.294(7)&$2^{30}$&0.455&0.533(3)\\
${\textbf{Run 2}}$&0.9&$2^{26.5}$&11.72(2)&$2^{30}$&0.436&0.515(2)\\
${\textbf{Run 3}}$&0.9&$2^{31.25}$&16.63(5)&$2^{32}$&0.415&0.493(3)\\
${\textbf{Run 4}}$&1.0&$2^{23.75}$&11.79(2)&$2^{28}$&0.512&0.422(2)\\
${\textbf{Run 5}}$&1.0&$2^{27.625}$&16.56(5)&$2^{32}$&0.498&0.400(1)\\
${\textbf{Run 6}}$&1.0&$2^{31.75}$&23.63(14)&$2^{34}$&0.484&0.386(4)\\
${\textbf{Run 7}}$&0.9&$2^{34}$&20.34(6)&$2^{34}$&0.401&0.481(3)\\
\end{tabular}
\caption{Parameters for each of our numerical simulations:
$\Tm,~\tw,~\xi(\tw)$, the longest simulation time $t_{\text {max}}$, the
replicon exponent $\theta$, and the value of $C_\mathrm{peak}(\tw)$ as defined
and employed in the zero-field-cooling protocol of
Ref.~\cite{zhai-janus:20a,zhai-janus:21}.
The replicon exponent $\theta$ is a function of $\tilde{x}= \ell_\mathrm{J}(T)/\xi(\tw)$, where $\ell_J(T)$ is the Josephson 
length~\cite{janus:18,zhai:19}.
}
\label{tab:details_num}
\end{ruledtabular}
\end{centering}
\end{table}
The $N=L^D$ Ising spins, $s_x=\pm 1$, interact with their lattice nearest neighbors through the Hamiltonian
\begin{equation}\label{eq:IEA_hamiltonian}
\mathcal{H}= -\sum_{\langle \boldsymbol x,\boldsymbol y\rangle}J_{\boldsymbol x\boldsymbol y}s_{\boldsymbol x} s_{\boldsymbol y} - H\sum_{\boldsymbol x} s_{\boldsymbol x},
\end{equation}
where the quenched disorder couplings are
$J_{\boldsymbol x \boldsymbol y}=\pm 1$, with 50 \% probability.  We name a
particular choice of the couplings a {\it sample}.  In the absence of an
external magnetic field ($H=0$), this model undergoes a spin-glass transition
at a critical temperature in simulation units 
$T_\text{g}=1.102(3)$~\cite{janus:13}. We study the off-equilibrium dynamics of the
model~\eqref{eq:IEA_hamiltonian} using a Metropolis algorithm (one lattice
sweep roughly corresponds to one picosecond of physical time).

We have studied a single sample (see Ref.~\cite{zhai-janus:20a,zhai-janus:21}
for sample variability studies). For each of the considered protocols, we have
developed 1024 statistically independent system trajectories (termed
\emph{replicas}), except for Runs 6 and 7 in Table \ref{tab:details_num} for
which we have simulated 512 replicas. Further simulation details can be found
in Table \ref{tab:details_num} (the rationale for our choices of temperatures
and magnetic fields is explained in Ref.~\cite{zhai-janus:20a}).

In order to follow the experimental protocols, the following procedures were taken:
\begin{itemize}
\item For the TRM protocol, the initial random spin configuration was placed instantaneously at the working temperature $\Tm$ in a magnetic field $H$ (direct quench).  It was allowed to relax for a time $\tw$ in the presence of $H$, after which the magnetic field was removed, and the magnetization,
\begin{equation}\label{eq:M_TRM_def}
M_\mathrm{TRM}(t,\tw;H)={\frac {1}{160^3}}\sum_{\boldsymbol x}s_{\boldsymbol x}(t+\tw;0),
\end{equation}
as well as the temporal auto-correlation function,
\begin{equation}\label{eq:C_TRM_def}
C_\mathrm{TRM}(t,\tw;H)={\frac {1}{160^3}}\sum_{\boldsymbol x}s_{\boldsymbol x}(\tw;H)s_{\boldsymbol x}(t+\tw;0),
\end{equation}
were recorded.
\item For the ZFC protocol, the initial random spin configuration was placed instantaneously at the working temperature $\Tm$ (direct quench) and allowed to relax for a time $\tw$ at $H=0$.  At time $\tw$, the magnetic field $H$ was applied and the magnetization,
\begin{equation}\label{eq:M_ZFC_def}
M_\mathrm{ZFC}(t,\tw;H)={\frac {1}{160^3}}\sum_{\boldsymbol x}s_{\boldsymbol x}(t+\tw;H),
\end{equation}
as well as the temporal auto-correlation function,
\begin{equation}\label{eq:C_ZFC_def}
C_\mathrm{ZFC}(t,\tw;H)={\frac {1}{160^3}}\sum_{\boldsymbol x}s_{\boldsymbol x}(\tw;0)s_{\boldsymbol x}(t+\tw;H),
\end{equation}
were computed.
\end{itemize}
Note that the auto-correlation function can be obtained experimentally as well in the TRM protocol~\cite{herisson:02}. Indeed, in the limit $H\to 0$, one has
$C_\mathrm{TRM}(t,\tw;H)\propto \langle M_\mathrm{TRM}(\tw)
M_\mathrm{TRM}(t+\tw)\rangle$, where
$\langle\ldots \rangle$ indicates the average over the thermal noise. Although
$\langle M_\mathrm{TRM}(\tw) M_\mathrm{TRM}(t+\tw)\rangle\propto
\sum_{{\boldsymbol x},{\boldsymbol y}}\,\langle s_{\boldsymbol x}(\tw;H)s_{\boldsymbol
  y}(t+\tw;0)\rangle$, the gauge invariance~\cite{toulouse:77} of the
Hamiltonian~\eqref{eq:IEA_hamiltonian}, that holds for $H\to 0$, ensures that only terms with
${\boldsymbol x}={\boldsymbol y}$ are non-vanishing in the double
sum.\footnote{The null contribution (in average) of the cross terms ${\boldsymbol x}\neq{\boldsymbol y}$ makes
  $\langle M_\mathrm{TRM}(\tw) M_\mathrm{TRM}(t+\tw)\rangle$ rather noisy, as
  it can be appreciated in Ref.~\cite{herisson:02}.}

In Appendix \ref{appendix:cooling_protocols} we discuss in detail the influence of using  different cooling protocols on observables measured in the numerical part of this paper (direct cooling). The conclusion of the Appendix is that we can use the direct quench protocol to confront with the experimental studies.

\subsection{The superposition principle breaks down for finite magnetic fields}
\label{subsec:violation_superpostion}

The experimental investigation of spin glasses is based on ZFC and TRM protocols, 
which are related to each other in the limit $H \to 0^+$ through Eq.~\eqref{eq:superposition_M}.

In this Section, we will show how the range of  validity for Eq.~\eqref{eq:superposition_M} works in numerical simulations which are designed to mimic the single crystal TRM/ZFC experiments (see Refs.~\citep{zhai-janus:20a,zhai-janus:21} for details about the lab/numerical simulation equivalence).

Let us rewrite the \emph{extended principle of superposition}, Eq.~\eqref{eq:superposition_M}, for simplicity as:
\begin{equation*}
M_{\text {TRM}}(t,\tw) + M_{\text {ZFC}}(t,\tw) = M_{\text {FC}}(0,\tw+t)\,.
\end{equation*}

\begin{figure}[h]
\centering
\includegraphics[width = 1\columnwidth]{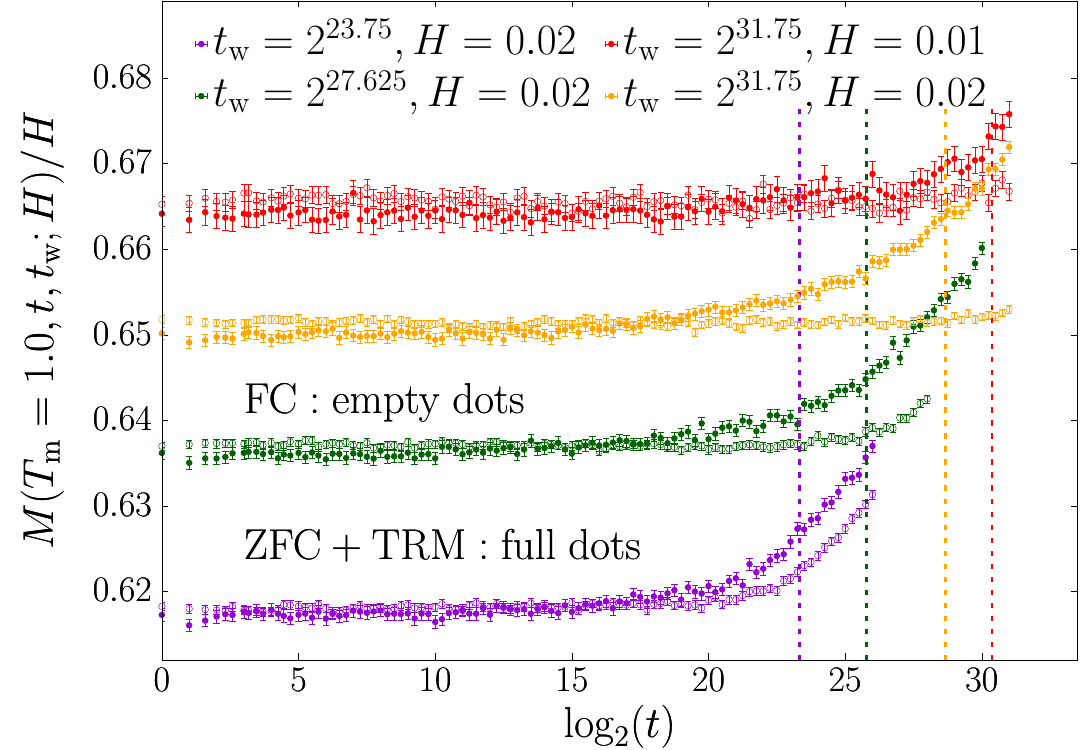}
\caption{Growth of the rescaled magnetizations for different experimental
protocols: $[M_\mathrm{ZFC}(t,\tw)+M_\mathrm{TRM}(t,\tw)]/H$, and
$M_\mathrm{FC}(0,\tw+t)/H$. The lighter colors are referred to the FC case; the darker ones are for the quantity $\mathrm{ZFC} + \mathrm{TRM}$. The vertical dashed lines correspond to the
effective times, $\teff_H$, associated with each case. The violation of
Eq.~\eqref{eq:superposition_M} is evident.}
\label{fig:superposition_regime_M_overH}
\end{figure}

We analyze separately the growth of the left-hand side,
$M_\mathrm{ZFC}(\tw,t)+M_\mathrm{TRM}(\tw,t)$, and the right-hand side,
$M_\mathrm{ZFC}(0,\tw+t)$, of the above expression. As the reader
notices in Fig.~\ref{fig:superposition_regime_M_overH}, when the magnetic field
increases, the violation of Eq.~\eqref{eq:superposition_M} increases; and the field-cooled magnetization, $M_\mathrm{FC}(0,\tw+t)$, changes with time. 

In order to characterize the violation of Eq.~\eqref{eq:superposition_M}, let us define the quantity
\begin{equation}
\begin{split}
D(t,\tw;H) \equiv & \frac{1}{H} \big(M_\mathrm{FC}(0,\tw+t)\\
& -\left[ M_\mathrm{ZFC}(t,\tw)+M_\mathrm{TRM}(t,\tw) \right]\big) \; .
\end{split}
\end{equation}
If the extended super position principle, Eq.~\eqref{eq:superposition_M}, is only valid for $H \to 0$, we can hypothesize that  $D(t,\tw;H)$ could behave as:
\begin{equation}\label{eq:superpostion_Delta_expantion}
D(t,\tw;H) = b_2(\tw;t) H^2 + b_4(\tw;t) H^4 + \mathcal{O}(H^6) \; ,
\end{equation}
where the coefficients $b_2(\tw;t)$ and $b_4(\tw;t)$ are some unknown functions
 and $\mathcal{O}(H^6)$ represents higher-order terms.

\begin{figure}[h]
\centering
\includegraphics[width = 1\columnwidth]{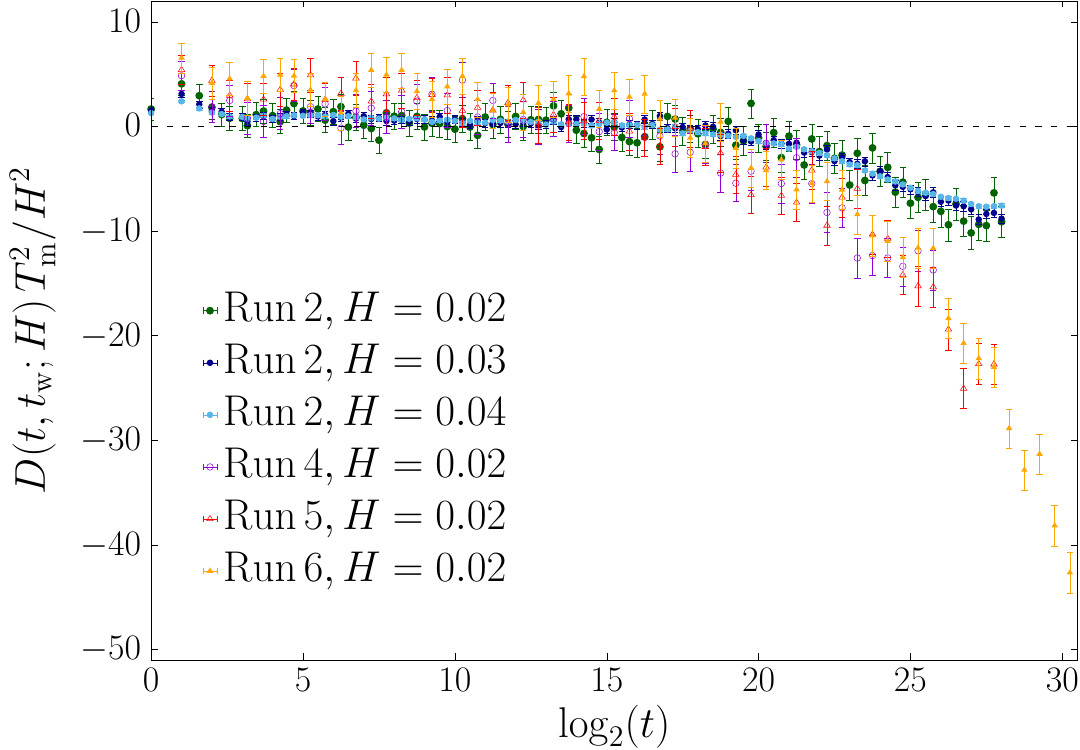}
\caption{Non-linear dependence of the rescaled quantity $D(t,\tw;H)\,\Tm^2/H^2$
  as a function of time. For clarity, we have omitted data at $H=0.01$ (errors
  are larger than 100\% for this field).}
\label{fig:superpostionDelta_rescaledHT}
\end{figure}

To test Eq.~\eqref{eq:superpostion_Delta_expantion}, we address the temporal behavior of the rescaled quantity, $D(t,\tw;H)\,\Tm^2/H^2$ \footnote{We have rescaled the quantity $D(t,\tw;H)$ by the temperature to compare data at different temperatures}, in Fig.~\ref{fig:superpostionDelta_rescaledHT}.

In Fig.~\ref{fig:superpostionDelta_rescaledHT}, we analyze two aspects:
\begin{itemize}
\item For a given waiting time $\tw$ [\emph{i.e.}, $\xi(\tw)$], what is the effect of increasing the magnetic field, $H$?
\item For a given value of the external magnetic field, $H=0.02$, what is the effect of changing the waiting time, $\tw$, [\emph{i.e.} $\xi(\tw)$]?
\end{itemize}

The answer to the first question is straightforward. From
Fig.~\ref{fig:superpostionDelta_rescaledHT}, increasing the magnetic field
$H$ causes the separation between the $H$ curves to increase as well. To the second
question, our understanding is that the violation of Eq.\eqref{eq:superposition_M} is caused by the difference between the time developments of $\xi_\mathrm{TRM}(t,\tw;H)$ and
$\xi_\mathrm{ZFC}(t,\tw;H)$. The lack of a dependence on $\tw$ in the
$b_2(\tw;t)$ and $b_4(\tw,t)$ coefficients is consistent with the expectation
that the only $\tw$ dependence lies within the $\tw$ dependence of the
correlation lengths themselves.  Otherwise, there would be a $\tw$ dependence
even in the $H^2 \to 0$ limit. 
In the next subsection, we will emphasize and display the difference between the TRM and ZFC protocols.

\section{Difference Between the ZFC and the TRM Protocols}
\label{sec:diff_TRM_ZFC_num}

One of the main differences between experiments and simulations is access to
the microscopic spin configurations.  Eq.~\eqref{eq:scaling_law_decay_teff}
could be read as a bridge to connect the macroscopic observable of the
effective time $\teff_H$ and the microscopic spin rearrangement
$\xi_\mathrm{micro}$.

Numerically, we have easy access to the spin configuration enabling us to calculate the microscopic correlation length $\xi_\mathrm{micro}(\tw;H)$ (see Appendix \ref{appendix:xi_micro_def} for details).
From the experimental point of view, Eq.~\eqref{eq:scaling_law_decay_teff} determines an \emph{effective correlation length}, $\xi_\mathrm{eff}(\tw;H^2 \rightarrow 0)$. See Secs.~\ref{sec:difference_xi_in_exp} above and \ref{subsec:ximicro_vs_xieff} below.

We shall follow two approaches to claim the same result: the two experimental
protocols are not equivalent, and the presence, or absence, of an external
magnetic field in the \textit{thermal history} of a spin glass is not
negligible.

On the one hand, we analyze the effect of the external magnetic field on the
microscopic correlation length $\xi_\mathrm{micro}$ (directly accessible in
simulations). 

On the other hand, we follow the same experimental approach, see
Sec.~\ref{sec:difference_xi_in_exp}, in order to evaluate the magnetic response through
the lens of the effective time $\teff_H$.

Finally, we conclude this section by showing the equivalence between the microscopic correlation length, $\xi_\mathrm{micro}(\tw)$, and the \textit{effective} correlation length
$\xi_\mathrm{eff}(\tw)$ (which is also experimentally measured) through Eq.~\eqref{eq:scaling_law_decay_teff}.

\subsection{\boldmath Numerical approach: the effect of an external magnetic field through the lens of the microscopical correlation length $\xi_\mathrm{micro}(\tw)$ }
\label{subsec:num_approachZFC_TRM_xi_micro}

In the zero-field-cooled protocol, the system is cooled to the working
temperature $\Tm$ in the absence of an external magnetic field, which is then
switched on after a waiting time $\tw$. Thus, by definition, the ZFC
protocol can be described by its microscopic correlation length,
$\xi^\mathrm{ZFC}_\mathrm{micro}(\tw)$.
The thermoremanent protocol, conversely,
brings the system to $\Tm$  in the presence of an external magnetic field. This
implies that each run has its own $\xi^\mathrm{TRM}_\mathrm{micro}(\tw;H)$
before $H$ is turned off.

\begin{figure}[h!]\centering
\centering
        \includegraphics[width=\columnwidth]{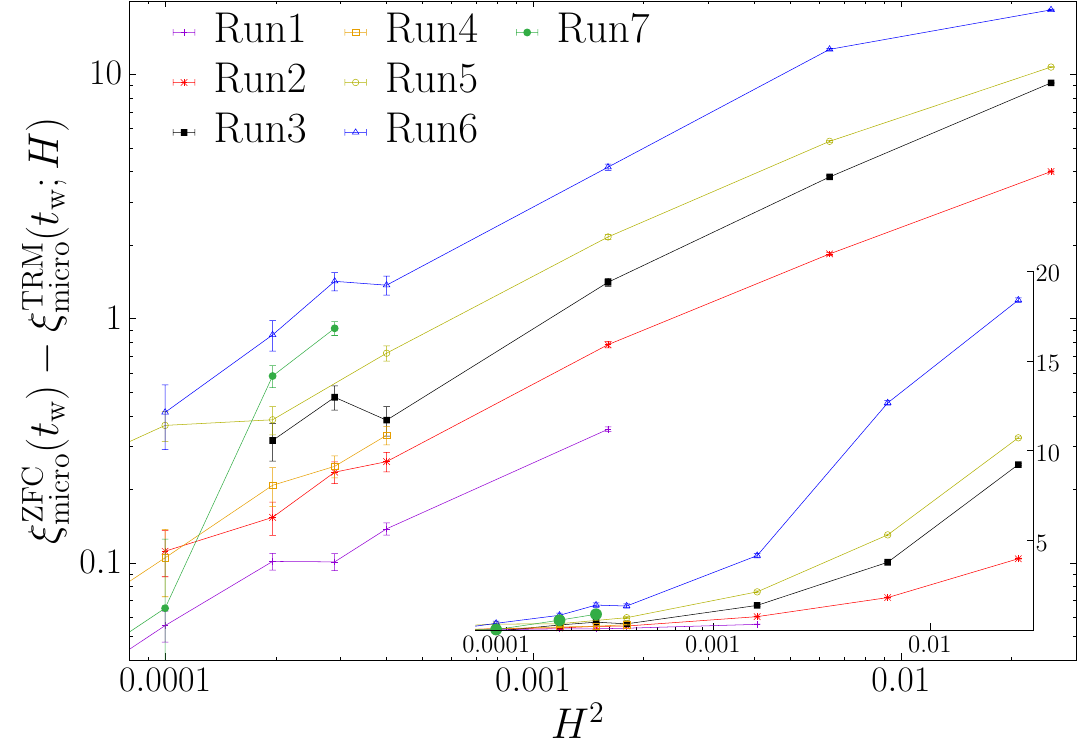}
  \caption{The difference between the ZFC and TRM  correlation length is plotted against $H^2$.
  (See Appendix ~\ref{appendix:xi_micro_def} for the definition of the microscopical correlation length).
  The \textbf{main} plot scale is log-log; the \textbf{insert} has a linear scale for
the ordinate. The ordering of the different runs (see Table
\ref{tab:details_num}) displays an increase of the difference with increasing
$\tw$.}
    \label{fig:diff_xi_micro_ZFC_TRM_vs_H2}
\end{figure}

Thus, in Fig.~\ref{fig:diff_xi_micro_ZFC_TRM_vs_H2} we display the difference between the ZFC and TRM  microscopic correlation length against $H^2$.

As can be seen from Fig.~\ref{fig:diff_xi_micro_ZFC_TRM_vs_H2}, the difference  $\xi^\mathrm{ZFC}_\mathrm{micro}(\tw;H)-\xi^\mathrm{TRM}_\mathrm{micro}(\tw;H)$ approaches the $H^2\rightarrow 0^+$
limit with a linear slope. The difference between the ZFC and TRM
correlation length appears to have the same behavior for different runs (i.e.
$\tw$).

Under equilibrium conditions, and for  large-enough correlation lengths $\xi_\mathrm{eq}$, there is a scaling theory between the correlation length and an external field $H$~\citep{parisi:88, amit:05}:
\begin{equation}
\label{eq:scaling_xi_vs_H} 
\xi(\tw) \propto H^{1/y_\mathrm{H}} , \quad  2 \, y_\mathrm{H}= D-\theta/2 
\end{equation}
where $D$ is the spatial dimension and $\theta$ is the replicon (see also Appendix 
\ref{appendix:scaling}).

\begin{figure}[h]
	\centering
	\includegraphics[width = 1\columnwidth]{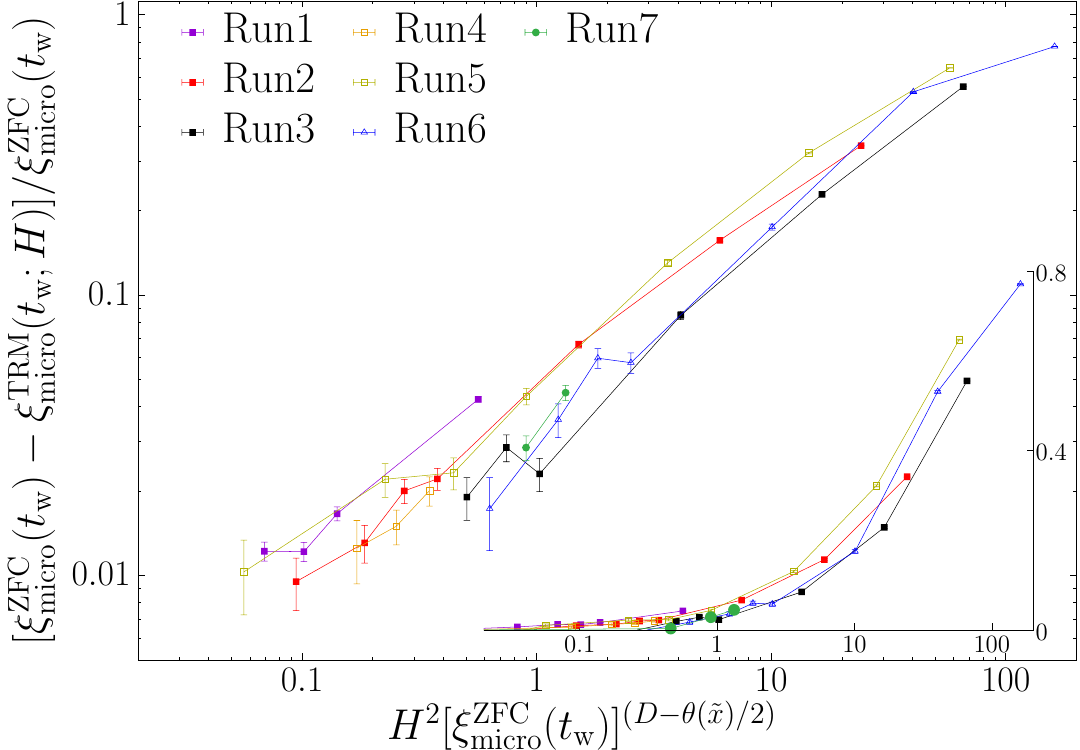}
	\caption{ The plot displays the dependence as a rescaled quantity $[\xi^\mathrm{ZFC}_\mathrm{micro}(\tw)-\xi^\mathrm{TRM}_\mathrm{micro}(\tw;H)]/\xi^\mathrm{ZFC}_\mathrm{micro}(\tw)$ against the function $H^2[\xi^\mathrm{ZFC}_\mathrm{micro}(\tw)]^{D-[\theta({\tilde x})/2]}$.  \textbf{The main plot} is in scale log-log; \textbf{the insert} has the linear scale for the ordinate.}
	\label{fig:naive_scaling_ximicro}
\end{figure}

This scaling behavior of $\xi(\tw)$ allows us to write the following rescaled relation for the two microscopic correlation lengths
\begin{equation}\label{eq:xi_naive_scaling}
\begin{split}
1- \xi^\mathrm{TRM}_\mathrm{micro}&(\tw;H)/ \xi^\mathrm{ZFC}_\mathrm{micro}(\tw)  \\
= & A(\tw,T)[\xi^\mathrm{ZFC}_\mathrm{micro}(\tw))]^{D-[\theta({\tilde x})/2]}H^2 \, .
\end{split}
\end{equation}
In Fig.~\ref{fig:naive_scaling_ximicro} we have tested that this scaling relation holds.

Finally remark that the logic behind the effect of the upward curvature of
$\varDelta(\tw)~\mathrm{vs}~\mathrm{Hd}$, as suggested in Ref.~\cite{lederman:91},
and discussed in Sec.~\ref{sec:dependence_Delta_vs_Hd_exp} of this paper,
requires that the difference $\xi_\mathrm{ZFC}(\tw)-\xi_\mathrm{TRM}(\tw;H)$
increases with increasing waiting time $\tw$.  This is because $\mathrm{Hd}$
itself increases with $\tw$, and hence the barrier height difference between
the ZFC and TRM protocols increases with $\tw$.

\subsection{Experimental approach: evaluation of the magnetic response through the effective times}
\label{subsec:exp_approachZFC_TRM_teff}

Now focus on the differences between the effective correlation lengths (TRM and ZFC) computed following the experimental approach.  
The effective waiting time, $\teff_H$, is fitted through Eq.~\eqref{eq:fitting_for_teff}. Let us rewrite this equation, for the TRM and ZFC protocols, as
\begin{equation}\label{eq:teff_ZFC_and_TR_explicit_dependence}
\begin{split}
&\log \teff_H(\mathrm{ZFC})=a_0^\mathrm{ZFC} +a_2^\mathrm{ZFC}(\tw;H) H^2 + \mathcal{O}(H^4)\\
&\log \teff_H(\mathrm{TRM})= a_0^\mathrm{TRM}+ a_2^\mathrm{TRM}(\tw;H)H^2 +\mathcal{O}(H^4)\, .
\end{split}
\end{equation} 

In order to isolate the correlation length contribution to the fitting coefficient $a_2(\tw;H)$, and exploiting as well the physical meaning of Eq.~\eqref{eq:scaling_law_decay_teff}, we  introduce the following notation:
\begin{equation} \label{eq:fitting_coefficient_K_xi_expression}
a_2(\tw,T) = - K\,\xi(\tw)^{D-(\theta/2)} \, ,
\end{equation}
where $\xi(\tw)$ stands for $\xi_\mathrm{micro}(\tw;H=0)$ and  $K$ is a \emph{generic} positive constant. As usual, we have dropped the temperature dependence of the replicon exponent, $\theta$.  Below, and in the next subsection, we shall give details about the  constant $K$.

By definition, the value of $\xi(\tw;H \to 0)$ is the same for both ZFC and TRM through the zero-fitting coefficient $a_0$. Thus, taking the difference between the two terms in Eq.~\eqref{eq:teff_ZFC_and_TR_explicit_dependence}, we obtain
\begin{equation}\label{eq:diff_log_teff_vs_xi}
\begin{split}
&\big[\log \teff_H(\mathrm{ZFC})-\log \teff_H(\mathrm{TRM}) \big]/\xi(\tw)^{D-(\theta/2)}\\
&=-\big[K^\mathrm{ZFC}-K^\mathrm{TRM}\big] H^2+
{\mathcal O}(H^4)\,.
\end{split}
\end{equation}
The above expression allows us to compare the two protocols directly, avoiding a precise determination for each of the coefficients,  $K^\mathrm{ZFC}$ and $K^\mathrm{TRM}$.  Moreover, Eq.~\eqref{eq:diff_log_teff_vs_xi}  shows that $\xi(\tw;H)$ differs between ZFC and TRM by terms of the order of $H^2$.

\begin{figure}[h]
	\centering
	\includegraphics[width = 1\columnwidth]{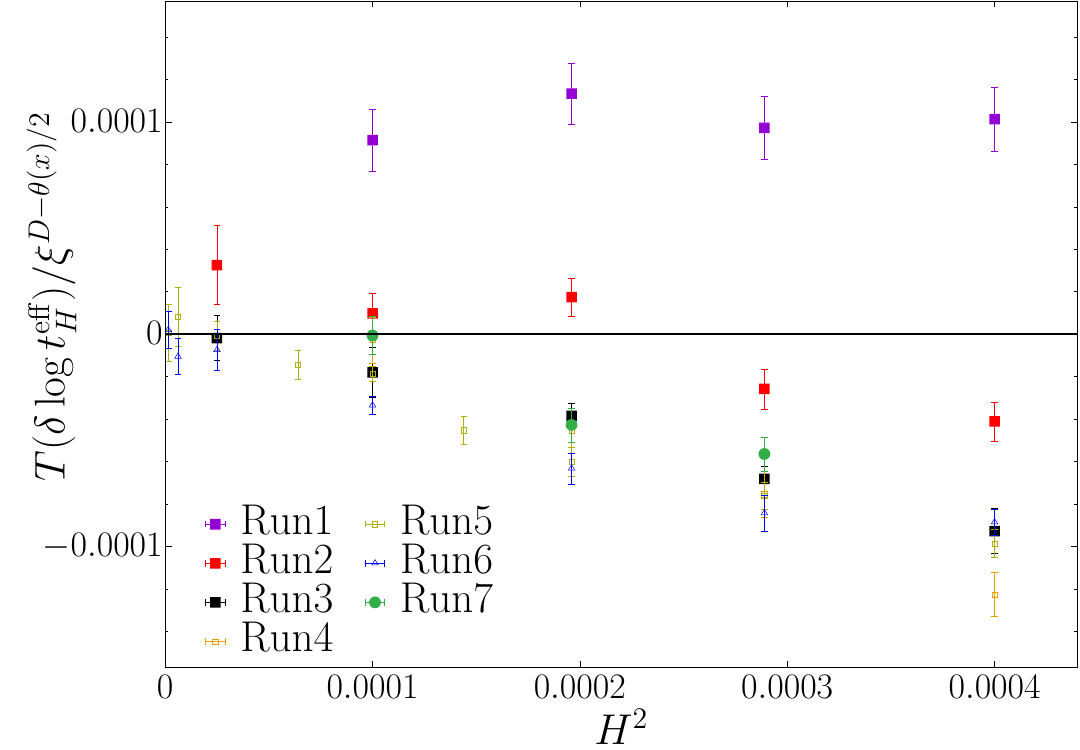}
	\caption{The rescaled quantity $T(\delta\log
\teff_H)/[\xi(\tw)]^{D-\theta/2}$ as a function of $H^2$.  The various run
details are listed in Table \ref{tab:details_num}. The correlation length
$\xi$ is the replicon correlation length $\xi_\mathrm{micro}(t,\tw;H=0)$, defined
through Eqs.~\eqref{eq:Gr_def}-\eqref{eq:xi_micro_def}.}
\label{fig:diff_log_teff}
\end{figure}

For ease of notation, we define,
\begin{equation}\label{eq:diff_log_teff_notation}
\log \teff_H(\mathrm{ZFC})-\log \teff_H(\mathrm{TRM})\equiv \delta\,\log \teff_H~~.
\end{equation}
We exhibit the rescaled quantity $T (\delta\,\log \teff_H)/[\xi(\tw)]^{D- \theta/2}$ as a function of $H^2$ in Fig.~\ref{fig:diff_log_teff}.

\begin{figure}[h]
	\centering
	\includegraphics[width = 1\columnwidth]{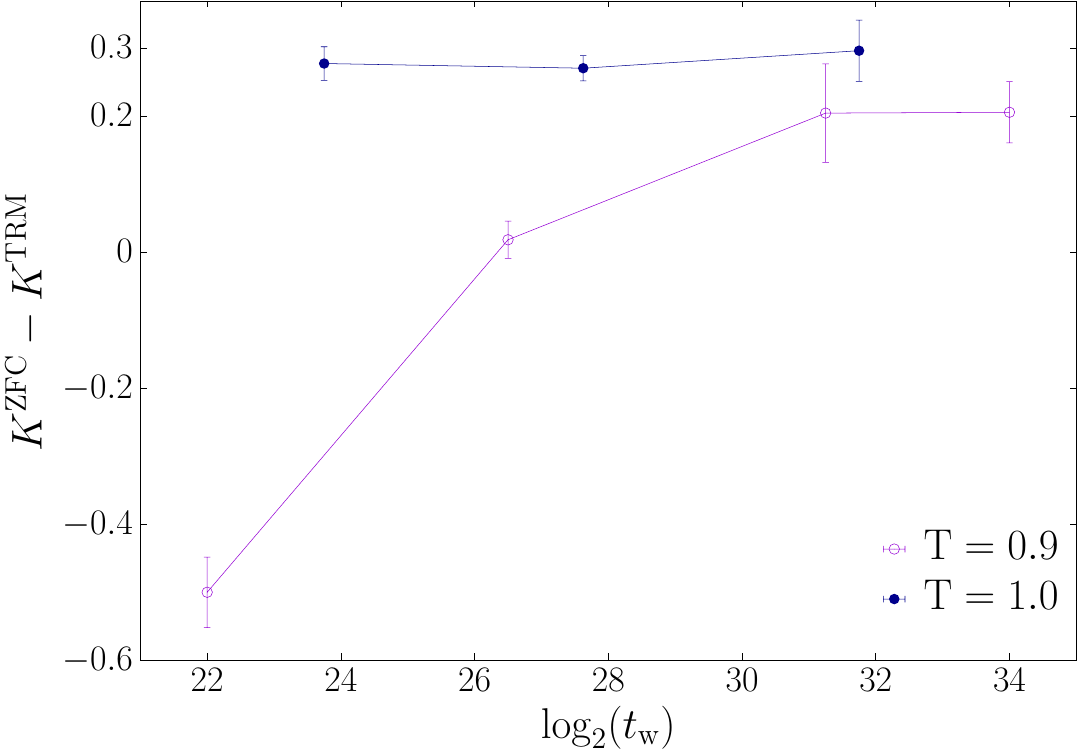}
	\caption{The difference between the ZFC and TRM decay of the slope of the ratio $\log (\teff_H/\teff_{H \to 0^+})$ as a function of the waiting time, $\tw$. By definition,  $K^\mathrm{ZFC}-K^\mathrm{TRM} = [a_2^\mathrm{ZFC}(\tw,T)-a_2^\mathrm{TRM}(\tw,T)]/\xi(\tw)^{D-\theta/2}$, see Eq.~\eqref{eq:fitting_coefficient_K_xi_expression}. In Tabs.~\ref{tab:ZFC_fit_num}-\ref{tab:TRM_fit_num} we report the fitting coefficients $a_2(\tw;T)$.}
	\label{fig:diff_K_ZFC_TRM_vs_tw}
\end{figure}

A scaling behavior for $\xi(\tw)$ sufficiently large is observed, as well as support for the principal relationship explored in this paper:
\begin{equation}\label{eq:comparison_Kzfc_Ktrm}
K^\mathrm{ZFC}> K^\mathrm{TRM}~~.
\end{equation}
This difference can be made quantitative by plotting the differences of the rescaled fitting coefficients, $\big(a_2^\mathrm{ZFC}(\tw,T)-a_2^\mathrm{TRM}(\tw,T)\big)/\xi(\tw)^{D-\theta/2}$  as a function of the waiting time, see Fig.~\ref{fig:diff_K_ZFC_TRM_vs_tw}. 

\subsection{\boldmath Comparison between the microscopic correlation length $\xi_{\text {micro}}(\tw;H)$ and the \textit{effective} correlation length $\xi_\mathrm{eff}(\tw;H)$}
\label{subsec:ximicro_vs_xieff}

We will study the relation between the microscopic correlation length (computed via the replicon propagator) and the effective one, $\xi_\mathrm{eff}(\tw)$, based on the fitting procedure introduced in Sec.~\ref{sec:difference_xi_in_exp}. The aim of this study is to test the relationship:
\begin{equation}\label{eq:xi_eff_is_xi_micro}
\xi_\mathrm{micro}(\tw;H=0) \simeq \xi_\mathrm{eff}(\tw;H^2 \rightarrow 0) \ .
\end{equation}

Let us begin with the ZFC case where the correlation length does not have any dependence on the external magnetic field. 

According to the scaling function Eq.~\eqref{eq:scaling_law_decay_teff}, the fitting coefficient $a_2^\mathrm{ZFC}(\tw;T)$, Eq.~\eqref{eq:fitting_for_teff}, behaves as
\begin{equation}\label{eq:c2_explicit_dependence_ZFC}
a_2^\mathrm{ZFC}(\tw;H) = \left[ \frac{\hat{S}}{2\Tm} \right] \xi(\tw)^{D-\theta(\hat{x})/2} \, .
\end{equation}
Therefore, taking into account Eq.~\eqref{eq:fitting_coefficient_K_xi_expression}, we can write:
\begin{equation}
\label{eq:K_ZFC_def}
K^\mathrm{ZFC}= -\frac{\hat{S}}{2\,\Tm} \,.
\end{equation}
We recall that this constant is positive.

Now, following the experimental procedure, see Sec.~\ref{sec:difference_xi_in_exp}, we can define an \emph{effective} correlation length by the ratio:
\begin{align}\label{eq:derivation_xi_eff_ZFC}
\frac{a_2^\mathrm{ZFC}(\tw,\Tm)}{a_2^\mathrm{ZFC}( \tw^*, \Tm)} = \frac{K^\mathrm{ZFC}(\tw)}{K^\mathrm{ZFC}(\tw^*)} \left[ \frac{\xi_\mathrm{eff}(\tw,\Tm)}{\xi_\mathrm{micro}(\tw^*;\Tm)} \right]^{D-\theta(\tilde{x})/2} \, ,
\end{align}
where $K^\mathrm{ZFC}(\tw) = K^\mathrm{ZFC}$, see Eq.~\eqref{eq:K_ZFC_def}, and we have omitted the $\xi(\tw)$ dependence of the replicon $\theta(\tilde{x})$.

Thus, we define $\xi_\mathrm{eff}^\mathrm{ZFC}(\tw,\Tm)$ as:
\begin{equation}\label{eq:xieff_def_num_ZFC}
\xi_\mathrm{eff}^\mathrm{ZFC} (\tw;\Tm) = \left[ \frac{a_2(\tw,\Tm)}{a_2( \tw^*, \Tm)}\right]^{1/(D- \theta(\tilde{x}) /2)} \hspace*{-.4cm}  \xi_\mathrm{micro}(\tw^*,\Tm) \, .
\end{equation}
The quantity $\xi_\mathrm{micro}(\tw^*,\Tm)$ plays the role of a reference length to avoid having to require a precise determination of the constants in Eq.~\eqref{eq:c2_explicit_dependence_ZFC}.

Let us now focus on the TRM protocol (see also Appendix~\ref{appendix:relaxation_S}).
Following the results of Sec.~\ref{subsec:exp_approachZFC_TRM_teff}, the equivalence between the TRM and ZFC protocols does not hold. We quantified the difference between the energetic barrier in the TRM and ZFC protocol in Fig.~\ref{fig:diff_K_ZFC_TRM_vs_tw}.
Let us formalize this difference as:
\begin{equation}\label{eq:formalization_diff_K}
K^\mathrm{ZFC}-K^\mathrm{TRM} = \tilde{B}(\tw) \; .
\end{equation}

By manipulating the above expression, we can obtain an expression for $K^\mathrm{TRM}$:
\begin{equation}\label{eq:K_TRM_expression}
K^\mathrm{TRM}(\tw)= K^\mathrm{ZFC}- \tilde{B}(\tw).
\end{equation}
Rewriting Eq.~\eqref{eq:xieff_def_num_ZFC} for the TRM case as:
\begin{align}\label{eq:derivation_xi_eff_TRM}
\frac{a_2^\mathrm{TRM}(\tw,\Tm)}{a_2^\mathrm{TRM}( \tw^*, \Tm)} = \left( \frac{K^\mathrm{TRM}(\tw)}{K^\mathrm{TRM}(\tw^*)} \right) \left[ \frac{\xi_\mathrm{eff}(\tw,\Tm)}{\xi_\mathrm{micro}(\tw^*;\Tm)} \right]^{D-\frac{\theta(\tilde{x})}{2}} \hspace*{-1.cm} ,
\end{align}
where $K^\mathrm{TRM}(\tw)$ has a weak dependence on the waiting time $\tw$ [ see Fig.~\ref{fig:diff_K_ZFC_TRM_vs_tw} and Eq.~\eqref{eq:K_TRM_expression}].\\
We obtain:
\begin{eqnarray}\label{eq:xieff_def_num_TRM_Istep}
\xi_\mathrm{eff}^\mathrm{TRM} (\tw;\Tm) &=& \left[ \frac{a_2^\mathrm{TRM}(\tw,\Tm)}{a_2^\mathrm{TRM}( \tw^*, \Tm)} \frac{K^\mathrm{TRM}(\tw^*)}{K^\mathrm{TRM}(\tw)} \right]^{\frac{2}{2D- \theta(\tilde{x})}} \nonumber\\ &\times&\xi_\mathrm{micro}(\tw^*,\Tm) \, .
\end{eqnarray}

\begin{figure}[h]
	\centering
	\includegraphics[width = 1\columnwidth]{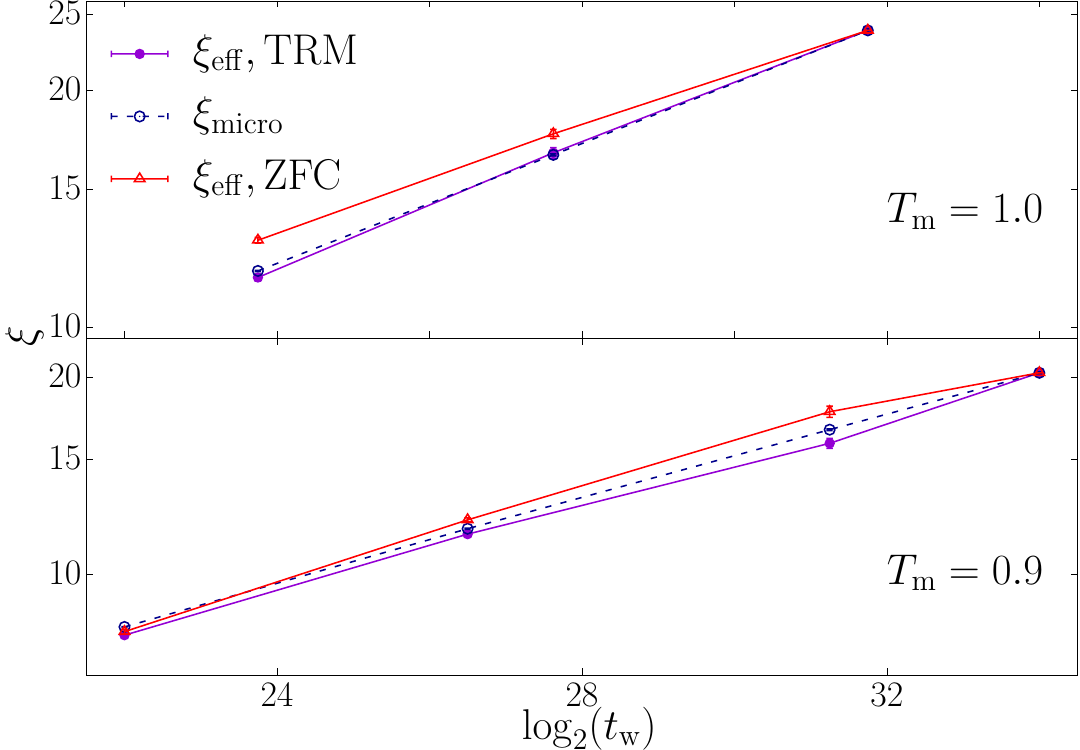}
	\caption{Comparison between $\xi_\mathrm{micro}(\tw;H\!=\!0)$ and $\xi_\mathrm{eff}(\tw;H^2 \!\rightarrow\! 0)$ as a function of the waiting time $\tw$. By definition of $\xi_\mathrm{eff}(\tw;H^2 \!\rightarrow\! 0)$, see Eqs.~\eqref{eq:xieff_def_num_ZFC} and \eqref{eq:xieff_def_num_TRM_Istep}, the $\tw^*$ taken as a reference has exactly the same $\xi_\mathrm{eff}(\tw^*,T) =\xi_\mathrm{micro}(\tw^*,T)$.} 
	\label{fig:xieff_ximicro_vs_tw}
\end{figure}

In Fig.~\ref{fig:xieff_ximicro_vs_tw}, we report the comparison between
$\xi_\mathrm{eff}(\tw;H)$ and $\xi_\mathrm{micro}(\tw;H)$ as a function of the
waiting time $\tw$ for both the ZFC and TRM cases. By definition, the $\tw^*$
taken as a reference has exactly the same $\xi_\mathrm{eff}(\tw^*,T)
=\xi_\mathrm{micro}(\tw^*,T)$.  We used as the reference $\tw^*=2^{34}$ at
$\Tm=0.9$ and $\tw^*=2^{31.75}$ at $\Tm=1.0$.

Given our lack of statistics, we could simulate only a single sample for each case. The errors are calculated from thermal fluctuations only. The numerical Ansatz of Eq.~\eqref{eq:xi_eff_is_xi_micro} is confirmed.

\section{Conclusion}
\label{sec:conclusion}

This paper provides a definitive basis for the observations of the Uppsala group regarding the generalized superposition principle. It present quantitative  evidence, both through experiments and numerical simulations, on how the superposition principle can be violated at finite magnetic fields.  The use of single crystals of CuMn enables experiments to exhibit the consequences of very large spin glass correlation lengths. The power of the Janus II supercomputer allows us to extend simulation times and length scales to values explored experimentally.   Both of these ingredients are vital for unveiling the difference in the magnetic response of the two experimental protocols that have been considered equivalent for more than three decades.

We have found that the nonlinear dependence of the energy barriers on the Hamming distance induces the breakdown of the generalized superposition principle.

Furthermore, the scaling law introduced in Refs.~\citep{zhai-janus:20a,zhai-janus:21}  has played the role of a touchstone for evaluation of the magnetic response of a 3D spin glass. In Sec.~\ref{subsec:ximicro_vs_xieff}
, we displayed the equivalence between the experimental extraction of the correlation length through Eq.~\eqref{eq:fitting_teff_num} and the microscopic calculation of $\xi$ through the replicon propagator ${\mathcal {G}}({\boldsymbol r}, \tw,T)$.

The unique and extraordinary collaboration between experiments, simulations, and theory has displayed once again its potential for the investigation of complex systems, as for the 3D spin glass.  We look forward to continued investigation of spin glass dynamics, building on the results of this paper as we examine the microscopic nature of such penomena as rejuvenation and memory.

\section*{Acknowledgements}
\addcontentsline{toc}{section}{Acknowledgements}
 We are grateful for helpful discussions with S. Swinnea about sample
characterization.  This work was supported in part by the U.S. Department of
Energy (DOE), Office of Science, Basic Energy Sciences, Materials Science and
Engineering Division, under Award No. DE-SC0013599, and performed at the Ames
Laboratory, which is operated for the U.S. DOE by Iowa State University under
contract No. DE-AC02-07CH11358.  We were partly funded as well by Ministerio de
Econom\'ia, Industria y Competitividad (MINECO, Spain), Agencia Estatal de
Investigaci\'on (AEI, Spain), and Fondo Europeo de Desarrollo Regional (FEDER,
EU) through Grants No. PID2021-125506NA-I00, No. PID2020-112936GB-I00, No. PID2019-103939RB-I00, No.PGC2018-094684-B-C21 and PGC2018-094684-B-C22, by the Junta de Extremadura
(Spain) and Fondo Europeo de Desarrollo Regional (FEDER, EU) through Grant No. 
GR21014 and No. IB20079  and by the DGA-FSE (Diputaci\'on General de Arag\'on --
Fondo Social Europeo).The research has received financial support from the Simons Foundation (grant No. 454949, G. Parisi) and ICSC – Italian Research Center on High Performance Computing, Big Data and Quantum Computing, funded by European Union – NextGenerationEU.
DY was supported by the Chan Zuckerberg Biohub and IGAP was supported by the Ministerio de Ciencia, Innovaci\'on y Universidades (MCIU, Spain) through FPU grant No. FPU18/02665.
B.S. acknowledges the support of the Agence Nationale de la Recherche (Ref. ANR CPJ-ESE 2022). JMG was supported by the Ministerio de Universidades and
the European Union NextGeneration EU/PRTR through 2021-2023 Margarita Salas
grant. I.P was a post-doctoral fellow at the Physics Department of Sapienza University of Rome during most part of this work.

\appendix
\section{Context for
  Sect.~\ref{sec:phenomenological_HD}}\label{appendix:phenomenological_HD}

From a phenomenological point of view, Eq.~\eqref{eq:superposition_M} depends upon the relationship between the barrier heights $\varDelta(t,\tw)$ between states and their \textit{separation}, termed the Hamming distance ($\mathrm{Hd}$), defined below in Eq.~\eqref{eq:Hdistance_def}.  If the relationship is linear, Eq.~\eqref{eq:superposition_M} holds.  If $\varDelta(t,\tw)$ increases more rapidly than linear with $\mathrm{Hd}$, Eq.~\eqref{eq:superposition_M}  will not hold for finite magnetic fields $H$. Under such conditions, the decay of the TRM will be slower than the rise of the ZFC, and the departure from Eq.~\eqref{eq:superposition_M}  will increase with increasing magnetic field change and increasing $\tw$.

In order to explain these dynamics, it is helpful to employ a simple phenomenological model that utilizes the concept of hierarchy of ancestors of spin-glass states.

One can organize the states, and their ancestors, according to ultrametric symmetry~\cite{mezard:84}.
Immediately after a temperature quench from above $T_\mathrm{g}$ to the measurement temperature $\Tm$, the spin-glass states have a ``self-overlap'' $q_{\alpha\alpha}(\Tm)\equiv q_{\text {EA}}(\Tm)$, the Edwards-Anderson order parameter~\citep{edwards:75}
\begin{equation}\label{eq:qEA_def}
q_{\alpha\alpha}(\Tm)\equiv q_{\text {EA}}(\Tm)={\frac
  {1}{N}}\sum_{\boldsymbol x} \langle s_{\boldsymbol x} (t=0)\rangle_\alpha^2 ~~,
\end{equation}
where $N$ is the number of (Ising) spins, $\boldsymbol x$ is the position
coordinate of spin $s$, and $\langle \cdots \rangle_\alpha$ represents a
thermal average restricted to a \emph{pure} state $\alpha$. Now, simple and
physically compelling as the above expression might look, we enter into a true
mathematical minefield by writing Eq.~\eqref{eq:qEA_def}. There are
difficulties of several kinds:
\begin{itemize}
  \item The decomposition of the Boltzmann weight into a sum over pure phases
    $\langle \cdots \rangle_\alpha$~\cite{ruelle:04} holds only in the
    thermodynamic limit.
  \item The random couplings make the process of taking the thermodynamic
    limit highly non-trivial (the construction of the metastate somewhat
    alleviates this problem~\cite{aizenman:90,newman:03,read:14,billoire:17,newman:22}).
  \item The state $\alpha$ in Eq.~\eqref{eq:qEA_def} is not an equilibrium
    state (as assumed in the two bullet points above), but a dynamic state. The
    characterization of the metastate in a dynamic context is only incipient~\cite{jensen:21}.
\end{itemize}
Similar caveats apply to Eq.~\eqref{eq:q_alpha_beta_time_evolution} below. In
fact, this Appendix should be read under the assumption that future
mathematical developments will make it possible to give a precise meaning to
expressions such as Eqs.~\eqref{eq:qEA_def}
and~\eqref{eq:q_alpha_beta_time_evolution}.

Nevertheless, the self-overlap $q_{\text {EA}}(T)$ can be computed without
recourse to the pure-state decomposition (see, \emph{e.g.}, Refs.~\cite{janus:10,janus:10b}).
One finds that, although $q_{\text {EA}}(T)$ is zero at $T=T_\mathrm{g}$, it rapidly
rises to unity as $T\rightarrow 0$. As the waiting time \tw grows,
the quenched initial state organizes from its
paramagnetic structure into a progressively growing spin glass state.  This is expressed
through a growing spin glass correlation length $\xi(t,\tw;H)$, which is a
function of the time elapsed from quench.  Here, $\tw$ is the time from quench
to when a change in magnetic field $H$ occurs, and the time $t$ begins just
after the change in magnetic field.
\begin{figure}[t]
	\includegraphics[width = 1\columnwidth]{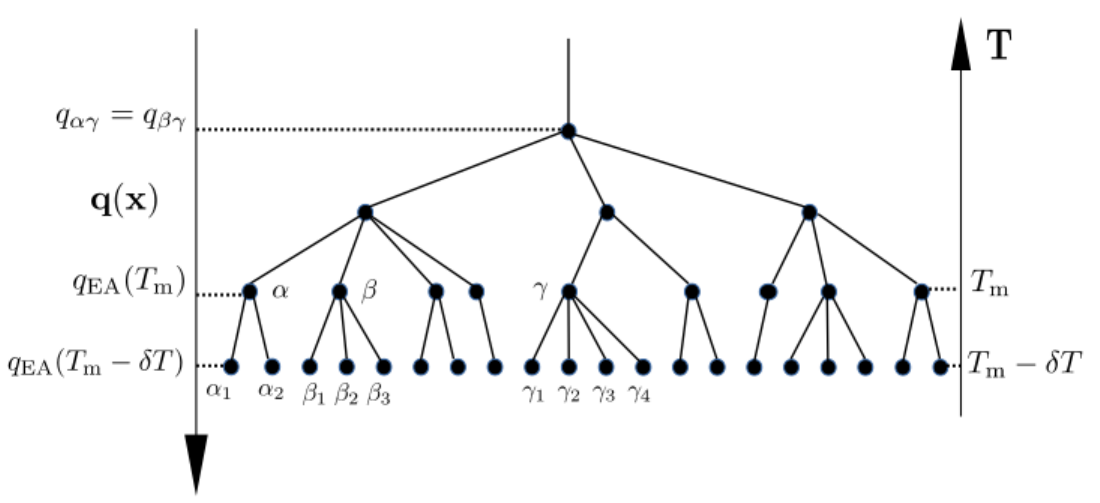}
	\caption{Three representative spin-glass states at the measurement
temperature $\Tm$ are labeled $\alpha,~\beta$ and $\gamma$.  Their overlaps
$q$, defined in Eq.~\eqref{eq:q_alpha_beta_time_evolution}, satisfy the
ultrametric topology $q_{\alpha\gamma}=q_{\beta\gamma}\leq q_{\alpha\beta}$.
The self-overlap, $q_{\alpha\alpha}=q_{\beta\beta}=q_{\gamma\gamma}=q_{\text
{EA}}(T)$, the Edwards-Anderson order parameter at temperature $T$.  Note that
as $q$ diminishes, the number of states increases exponentially.  When the
temperature is lowered from the quenched measuring temperature, $\Tm$ to
$\Tm-\delta T$, the states ``foliate'' to daughter states, each of which now
possesses the self-overlap $q_{\text {EA}}(\Tm-\delta T)$.  The states  are designated
by the symbols $\alpha_1,~\alpha_2$, etc. This figure is inspired by Fig. 9 of Ref.~\citep{parisi:06b}.}
	\label{fig:tree_daughter_random}
\end{figure}

As $\xi(t,\tw;H)$ grows, the overlap with the initial state $q_{\alpha\beta}$ diminishes:
\begin{equation}\label{eq:q_alpha_beta_time_evolution}
q_{\alpha\beta}={\frac {1}{N}}\sum_{\boldsymbol x} \langle s_{\boldsymbol x}^\alpha(t=0)s_{\boldsymbol x}^\beta(t,\tw;H) \rangle
\end{equation}
where the state has evolved from its initial quenched state $\alpha(t=0,\tw=0;H)$ to $\beta(t,\tw;H)$.  This is pictured in Fig.~\ref{fig:tree_daughter_random} for a random ultrametric tree~\citep{parisi:06b}.

\begin{figure}[h]
	\centering
		\includegraphics[width = 1\columnwidth]{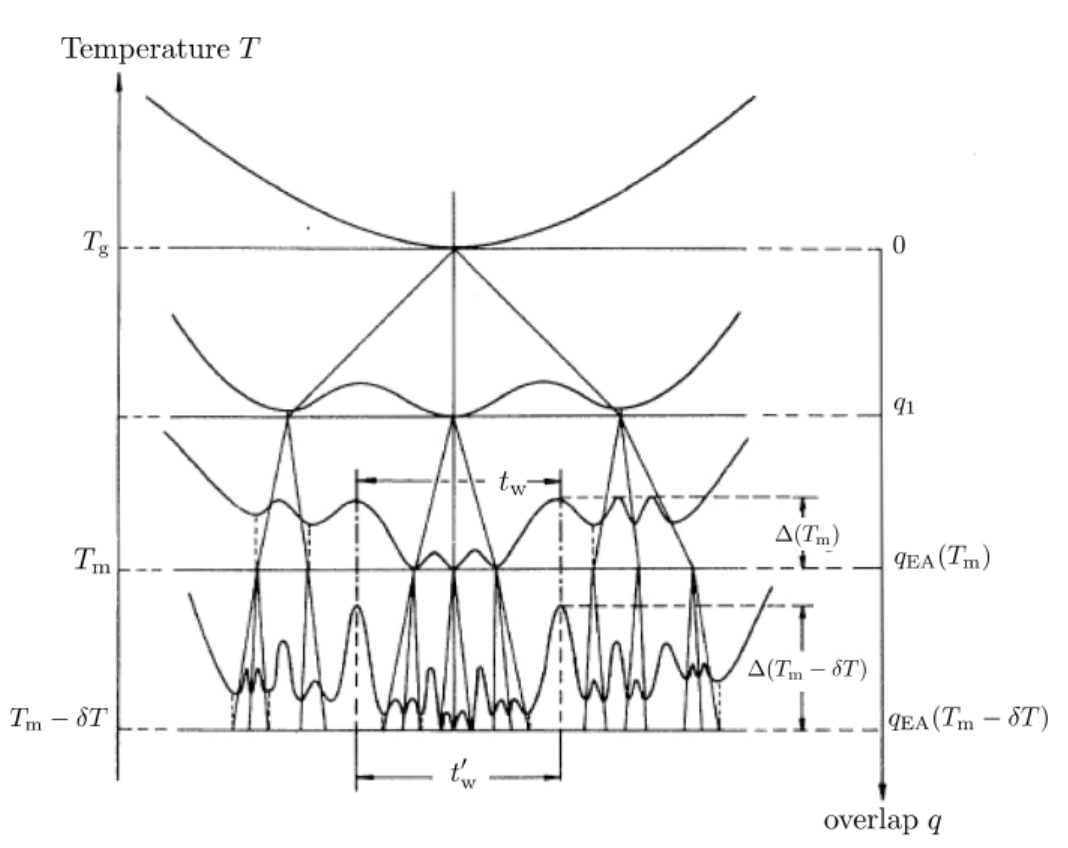}
	\caption{Hierarchical organization of metastable states.  The coarse-grained free-energy surface is represented at each level corresponding to a given temperature.  When the temperature is decreased, each valley subdivides into others.  The times $\tw$ and $t_{w}^\prime$ which are necessary to explore, at $\Tm$ and $\Tm - \delta T$, respectively, the region of phase space bounded by the same barriers are indicated.  The closest common ancestor to all states within the space bounded by $\varDelta_1$ at $\Tm$ and $\Tm-\delta T$ is the same, and its corresponding value of the overlap function is $q_1$.  The sketch also shows that, as the system explores more of phase space, it encounters ever increasing barrier heights, and that the free-energy surface has a self-similar structure. This figure is taken from Ref.~\cite{lederman:91}.}
	\label{fig:hierarchical_metastable_state}
\end{figure}

The ultrametric structure in Fig.~\ref{fig:tree_daughter_random} was
originally derived for Parisi's \textit{pure states}, a small proportion of the
total states available to the spin glass system.  The barriers between states
were infinite, so that no dynamics were present.  Experimentally, as developed
in Refs.~\cite{joh:96,joh:99},
it was not
only convenient, but quantitative, to include states with finite barriers obeying the same geometry
{\em between} the pure states. In fact, experiments over
a limited temperature range were shown to display a temperature dependence of
barrier heights that could be extrapolated to the pure state 
limit~\cite{hammann:92}.  A representative temperature-dependent organization of
metastable states is exhibited in Fig.~\ref{fig:hierarchical_metastable_state}.

We need some way to account for the distance between the states exhibited in Fig.~\ref{fig:tree_daughter_random}.  We express a Hamming distance, $\mathrm{Hd}$, in terms of the departure of the overlap from $q_\mathrm{EA}$ to $q_{\alpha\beta}$.  By this construction, the Hamming distance, $\mathrm{Hd}$, is
\begin{equation}\label{eq:Hdistance_def}
\mathrm{Hd} ={\frac {1}{2}}(q_\mathrm{EA}-q_{\alpha\beta})~~.
\end{equation}

One can intuitively think of a free-energy barrier separating states of
different $q_{\alpha\beta}$ as proportional to a \textit{distance} between
them, as shown in Fig.~\ref{fig:tree_daughter_random}.  That is, the barrier
heights should increase as some function of decreasing $q_{\alpha\beta}$ or
increasing $\mathrm{Hd}$.  A relationship between $\mathrm{Hd}$ and barrier
heights was first developed by Vertechi and Virasoro~\cite{vertechi:89}. 
In  Fig. 1 of Ref.~\cite{vertechi:89} it is shown that there is an upward curvature in the relationship between $\varDelta$ and $\mathrm{Hd}$ that we shall argue below can be extracted from experiment, and is exhibited in our simulations.

\begin{figure}[h]
	\centering
	\includegraphics[width = 1\columnwidth]{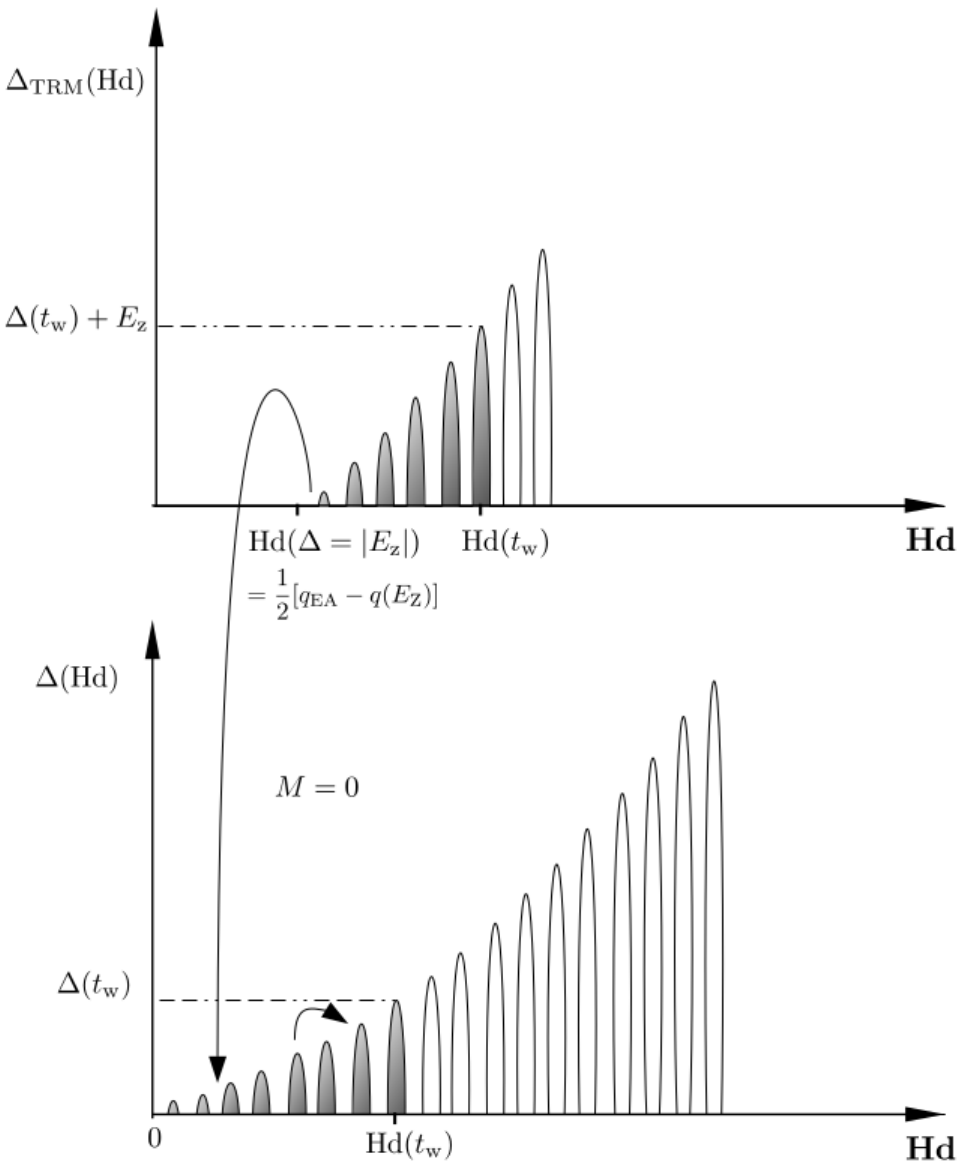}
	\caption{A cartoon illustrating the dynamics associated with the reduction of the barrier heights $\varDelta(t,\tw;H)$ by the ``Zeeman energy'' $E_\mathrm{Z}$~\cite{kenning:95}. The upper figure represents the system with magnetization $M_\mathrm{FC}$, the field-cooled magnetization, while the lower figure represents the system at $H=0$ with magnetization $M=0$. See the text for further explanation.}
	\label{fig:cartoon_dynamics_Delta}
\end{figure}

Consider now the Zeeman energy $E_\mathrm{Z}$ in
Eq.~\eqref{eq:Delta_reduction_Ez}. Its effect on barrier heights is treated
equivalently by Bouchaud et \textit{al}.~\cite{vincent:95} and Joh et
\textit{al}.~\cite{joh:99} in terms of a trap model and a barrier model,
respectively.  One can visualize the effect on the barrier heights using the
pictorial description exhibited in Fig.~\ref{fig:cartoon_dynamics_Delta} as
developed in Ref.~\cite{kenning:95}.

The figure is meant to illustrate the dynamics associated with a TRM protocol.
A magnetic field $H$ is applied to the spin glass at a temperature above the
condensation temperature $T_\mathrm{g}$ (or critical one).  Without changing $H$, the system is
cooled to the working temperature $\Tm <T_\mathrm{g}$.  The lower part of the free energy of
the system is then the upper part of Fig.~\ref{fig:cartoon_dynamics_Delta}.
All of the states in Fig.~\ref{fig:cartoon_dynamics_Delta} between the
barriers are assumed to have the same magnetization. This assumption is consistent with the near constancy in time of the field-cooled magnetization in experiments, 
$M_\mathrm{FC}(0,\tw+t)$ (see Eq. (\ref{eq:superposition_M}))~\cite{djurberg:99}.  
The system is allowed to \emph{age} at $\Tm$ for a time
$\tw$ in the presence of $H$.  This results in the growth of the correlation
length $\xi(t=0,\tw;H)$ creating ever increasing barriers up to the value $\varDelta(t=0,\tw;H)$ at the Hamming distance $\mathrm{Hd}(\tw)$. 

The effect of the magnetic field is expressed through the Zeeman energy
$E_\mathrm{Z}$, Eq.~\eqref{eq:zeeman_energy_def}.  Both models (trap/barrier)
reduce the depth/height of all of the traps/barriers by the same amount,
$E_\mathrm{Z}$. In the barrier model, those barriers for which
$\varDelta<|E_\mathrm{Z}|$ are assumed to vanish, so that the growth of the
correlation length begins at an $\mathrm{Hd}$ set by $\varDelta=|E_\mathrm{Z}|$.
That means that $\mathrm{Hd}(\tw; H)$ is larger than it would be at
$\mathrm{Hd}(\tw; H=0)$. This point will be crucial when we compare
$\xi(t,\tw;H)$ for TRM experiments with ZFC experiments.

In a TRM experiment, after aging for $\tw$ in the presence of $H$, the field
is cut to zero and the remanent magnetization $M_\mathrm{TRM}(t,\tw;H)$
measured.  By construction in Fig.~\ref{fig:cartoon_dynamics_Delta}, the
states with magnetization $M_\mathrm{FC}$ now are higher in energy than for
those with $M=0$.  There is an instantaneous decay of those states in the
$M_\mathrm{FC}$ manifold for which $\varDelta$ is less than $E_\mathrm{Z}$ to the
ground $M=0$ manifold (the ``reversible magnetization''
change)~\cite{kenning:95}.  The measurement time $t$ begins when $H$ is cut to
zero.  Diffusion from the remaining occupied states to the ground state
($M=0$) reduces the magnetization in the (now) higher energy manifold as shown
by the arrow in Fig.~\ref{fig:cartoon_dynamics_Delta}. The exponentially
increasing degeneracy exhibited in Fig.~\ref{fig:tree_daughter_random} leads
to the states with the highest barrier dominating the decay.

A consequence of Fig.~\ref{fig:cartoon_dynamics_Delta} is that, in the presence of a magnetic field $H$, the barriers are quenched for values of $\mathrm{Hd}$ such that $\varDelta(\mathrm{Hd})\leq |E_\mathrm{Z}| \equiv \varDelta_{E_\mathrm{Z}}(\mathrm{Hd})$ with $E_\mathrm{Z}$ defined by Eq.~\eqref{eq:zeeman_energy_def}. Eq.~\eqref{eq:Hdistance_def} can be written alternatively to take into account this cancellation.  Take $q_{\alpha\beta}(E_\mathrm{Z})=q_\mathrm{EA}-q(E_\mathrm{Z})$ to be the value of $q_{\alpha\beta}$ at the value of $q$ where $\varDelta(q)=|E_\mathrm{Z}|$.  Then, from Eq.~\eqref{eq:Hdistance_def}, the $\mathrm{Hd}$ has the value,
\begin{equation}\label{eq:Hd_vs_Ez}
\mathrm{Hd}(|E_\mathrm{Z}|=\varDelta)={\frac {1}{2}}[q_\mathrm{EA}-q(E_\mathrm{Z})]
\end{equation}
at the value of $q$ for which $|E_\mathrm{Z}|=\varDelta(q)$. The probability of
finding a value of $q$ larger than $q(E_\mathrm{Z})$ is zero. That is, there are no values of $q$ larger than this value.

As a consequence, $\xi(t=0,\tw;H)$ grows from zero not from $q=0$ but rather from $q=q(E_\mathrm{Z})$ in a TRM experiment.  If $\varDelta(\tw)$ would depend linearly on $\mathrm{Hd}$, equivalently on $q_{\alpha \beta}$ from Eq.~\eqref{eq:Hdistance_def}, there would be no difference in $\xi(t=0,\tw;H)$ between a TRM and a ZFC experiment.
In both, the growth of $\xi(t=0,\tw;H)$ would depend upon $\mathrm{Hd}$ in the same manner.  However, if instead $\varDelta(t=0,\tw;H)$ behaves as drawn in Fig.~\ref{fig:cartoon_dynamics_Delta}, that is, experiences an upward curvature as $\mathrm{Hd}$ increases, then beginning at a larger value of $q$ in a TRM protocol would mean that the growth of $\varDelta(t=0,\tw;H)$ would be slower for a TRM protocol than for a ZFC protocol.

Lederman \emph{et al.}~\cite{lederman:91} found evidence for an upward curvature of $\varDelta$ as a function of $\mathrm{Hd}$ in a $\mathrm{AgMn} \, 2.6$ at.\% spin glass.  Using measurements of the temperature dependence of both $\varDelta$ and $q_\mathrm{EA}$, they extracted the ratio,
\begin{equation}\label{eq:variation_Delta_Hd}
{\frac {\delta \varDelta}{\delta \mathrm{Hd}}}\bigg|_T=\alpha(T)\varDelta-\beta(T)\,,
\end{equation}
where $\alpha(T)$ and $\beta(T)$ are positive constants dependent only on temperature.  Integration of Eq.~\eqref{eq:variation_Delta_Hd} gives,
\begin{equation}\label{eq:variation_Delta_vs_Hd}
\varDelta(\mathrm{Hd})= C e^{\alpha(T) \mathrm{Hd}}+{\frac {\beta(T)}{\alpha(T)}}\,.
\end{equation}
where $C$ is an integration constant.
Thus, as $H$ increases, the diffusion in the time $\tw$ encounters ever increasing barrier heights for TRM protocols as compared to ZFC protocols. This is the
basis for our analysis that
$\xi_{\text{ZFC}}(t,\tw;H)>\xi_{\text{TRM}}(t,\tw;H)$ in the text.

Despite the tree based picture provides a good qualitative and even quantitative description of spin glass experiments~\cite{vincent:22}, we will finish this section discussing theories that differ from RSB in their predicted structure for the spin glass phase: the droplet model (DM), the Trivial-Nontrivial theory (TNT) and the Chaotic pairs (CP) picture.

The droplet model states that the spin glass phase is composed of only two pure states (related by the inversion symmetry). Certainly, the two states of the DM, when complemented with the notion of temperature chaos, are succesful in describing  the weak cooling-rate dependence found in experiments (and  in numerical simulations, see Appendix \ref{appendix:cooling_protocols}). However,  other interesting experiments such as the memory and rejuvenation effects~\cite{jonason:98}, are difficult to describe within the DM framework. Indeed, as discussed in Ref.~\cite{jonason:98}, a more complicated structure of fractal domains inside domains is the least that one needs to account for memory and rejuvenation (RSB' space filling excitations would be an extreme example of the "fractal within domain'' picture).

Furthermore,  the replicon exponent, that is used in practically all the analysis performed in this paper, is zero in the DM, and the Hamming distance is trivial (because it takes only two values). Finally, let us recall that in the DM the dependence of the dynamic correlation length as a function of time is given by $\xi(t_w) \sim [\log t_w] ^{1/\psi}$, where $\psi$ is the droplet exponent, controlling the free-energy barriers in the dynamics, instead of the power law of time found in experiments and numerical simulations~\cite{janus:18}. Besides, an extrapolation of Janus results to the experimental length scale assuming the droplets scaling found an aging rate much larger than experimentally found~\cite{janus:18}.

As we said above, there are two other pictures of the spin glass phase, namely TNT and Chaotic Pairs. TNT has no predictions regarding the out of equilibrium behavior of spin glasses: it only states that the equilibrium  probability distribution of the overlap has the shape predicted by RSB but the link overlap follows the DM prediction (it is trivial) \cite{krzakala:00,palassini:00}. As for the Chaotic Pairs picture, it  was introduced in the (equilibrium) mathematical physics literature as a theory with a dispersed  metastate (as RSB). In fact, the CP picture has not been developed outside the rigorous studies of spin glasses in equilibrium.\cite{read:14}

\section{Results from another single
  crystal}\label{appendix:second-single-crystal}

In addition to the 8 at.\% CuMn single-crystal sample with $\Tm = 37.5$ K ($\Tg = 41.5$ K), the properties of which are reported in the text, a 6 at.\% CuMn single crystal sample was measured at $\Tm = 26$ K ($\Tg = 31.5$ K).  The lower measuring temperature resulted in smaller values of $\xi(t,\tw;H)$ because of the attendant slow growth rate, leading to a smaller difference in the difference between $\xi_{\text {ZFC}}(\tw;H)$ and $\xi_{\text {TRM}}(\tw;H)$ than for the $\Tm = 37.5$ K measurements.  Nevertheless, at the largest waiting time, $\tw = 10000$~s, the difference lies well outside of the error bars.

\begin{figure}[h]
	\centering
	\includegraphics[width = 1\columnwidth]{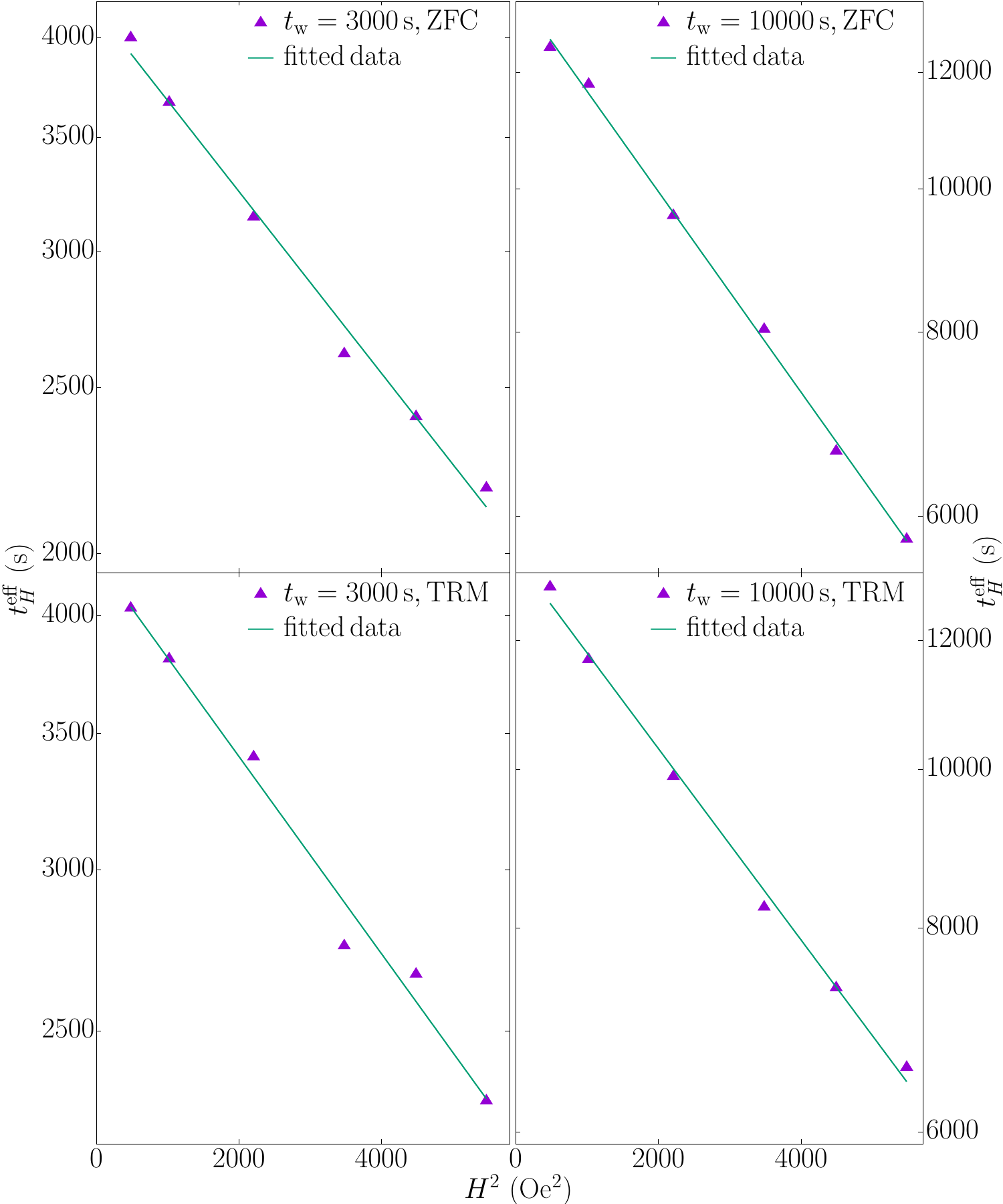}
	\caption{A plot of data and fit for the log of the effective waiting time, $\tw^{\text {eff}}$ vs $H^2$ for ZFC and TRM measurements on a 6 at.\% single crystal at $\tw=3000$ and $10000$~s and $T=26$ K ($\Tg = 31.5$ K).  The polynomial fitting parameters for each of the values of $H$, up to and including $H^4$, are given in Tables~\ref{tabappendixI}-\ref{tabappendixIV}.  The value of the correlation length is extracted from the $H^2$ fitting terms.}
	\label{figappendix:log_teff_T26_exp}
\end{figure}

The data are exhibited in Fig.~\ref{figappendix:log_teff_T26_exp} for the two values of the waiting time, $\tw$, at $T = 26$ K: $\tw = 3000$~s and $\tw = 10000$~s, for both the ZFC and TRM protocols.

From Fig.~\ref{figappendix:log_teff_T26_exp},
\begin{align*}
&\xi_{\text {ZFC}}(\tw\!=\!3000)-\xi_{\text {TRM}}(\tw\!=\!3000)=3.8(4.8) \\
&\xi_{\text {ZFC}}(\tw\!=\!10000)-\xi_{\text{TRM}}(\tw\!=\!10000)=7.7(2.6)
\end{align*}
taking the sum of the extreme limits of the error bars.
Assuming that $\varDelta(\tw)$
increases faster than linearly with Hd, the larger $\tw$, the slower
the growth of $\xi_{\text {TRM}}(\tw;H)$, as can be inferred from the
shape of $\varDelta$ vs Hd exhibited in
Fig.~\ref{fig:cartoon_dynamics_Delta} of
Appendix~\ref{appendix:phenomenological_HD}.  Thus, the difference
$\xi_{\text {ZFC}}(\tw)-\xi_{\text {TRM}}(\tw)$ should increase with
increasing $\tw$.  The data from
Fig.~\ref{figappendix:log_teff_T26_exp} in this Appendix
support this prediction: the difference $\xi_{\text
  {ZFC}}(\tw=10000)-\xi_{\text {TRM}}(\tw=10000)$ is much larger
than $\xi_{\text {ZFC}}(\tw=3000)-\xi_{\text {TRM}}(\tw=3000)$, and
well beyond the limits of the error bars.

Thus, the measurements on the 6 at.\% CuMn single crystal at 26 K provide additional experimental evidence for the assertion that $\xi_{\text {ZFC}}(\tw)$ is larger than $\xi_{\text {TRM}}(\tw)$ to that exhibited in the main text for an 8 at.\% CuMn single crystal.

The tables analogous to Tables~\ref{tab:zfc375t0}---\ref{tab:trm375}
for the 6 at.\% single crystal sample with $\Tm = 26$ K, are listed in
Tables~\ref{tabappendixI}---\ref{tabappendixIV}

\begin{table}[h]
	\begin{centering}
		\begin{ruledtabular}
			\begin{tabular}{c c c}
				$H$&$E_0$ & $E_2$ \\ 
				22.0&30.85298&$-0.04875$\\
				32.0&30.85298&$-0.10313$\\
				47.0&30.85298&$-0.22249$\\
				59.0&30.85298&$-0.35060$\\
				67.0&30.85298&$-0.45212$\\
				74.0&30.85298&$-0.55153$\\
			\end{tabular}
			\caption{Converted energy scale for ZFC protocol at $T=26$ K with $\tw = 3000$ s for various values of $H$ expressed in Oersteds.  The $E_n$ are the magnitude of the nth fitting parameter (including the respective $H^n$) expressed in units of $k_\mathrm{B} T_\mathrm{g}$ (see Eq. \eqref{eq:defEn}).  The correlation length $\xi_\mathrm{ZFC}=123(2)$ is derived from Eq.~\eqref{eq:xi_Zeeman_through_teff}. }\label{tabappendixI}
		\end{ruledtabular}
	\end{centering}
\end{table}
\begin{table}[h]
	\begin{centering}
		\begin{ruledtabular}
			\begin{tabular}{c c c}
				$H$&$E_0$ & $E_2$ \\ 
				22.0&30.87475&$-0.04457$\\
				32.0&30.87475&$-0.09431$\\
				47.0&30.87475&$-0.20344$\\
				59.0&30.87475&$-0.32059$\\
				67.0&30.87475&$-0.41342$\\
				74.0&30.87475&$-0.50432$\\
			\end{tabular}
			\caption{Converted energy scale for TRM protocol at $\Tm=26$ K with $\tw = 3000$ s for various values of $H$ expressed in Oersteds.  The $E_n$ are the magnitude of the nth fitting parameter (including the respective $H^n$) expressed in units of $k_\mathrm{B} T_\mathrm{g}$ (see Eq. \eqref{eq:defEn}).  The correlation length $\xi_\mathrm{TRM}=120(3)$ is derived from Eq.~\eqref{eq:xi_Zeeman_through_teff}.}\label{tabappendixII}
		\end{ruledtabular}
	\end{centering}
\end{table}
\begin{table}[h]
	\begin{centering}
		\begin{ruledtabular}
			\begin{tabular}{c c c}
				$H$&$E_0$ & $E_2$ \\ 
				22.0&31.83377&$-0.06386$\\
				32.0&31.83377&$-0.13512$\\
				47.0&31.83377&$-0.29148$\\
				59.0&31.83377&$-0.45932$\\
				67.0&31.83377&$-0.59233$\\
				74.0&31.83377&$-0.72256$\\
			\end{tabular}
			\caption{Converted energy scale for ZFC protocol at $\Tm=26$ K with $\tw = 10000$ s for various values of $H$ expressed in Oersteds.  The $E_n$ are the magnitude of the nth fitting parameter (including the respective $H^n$) expressed in units of $k_\mathrm{B} T_\mathrm{g}$ (see Eq. \eqref{eq:defEn}).  The correlation length $\xi_\mathrm{ZFC}=135(1)$ is derived from Eq.~\eqref{eq:xi_Zeeman_through_teff}.}\label{tabappendixIII}
		\end{ruledtabular}
	\end{centering}
\end{table}
\begin{table}[h]
	\begin{centering}
		\begin{ruledtabular}
			\begin{tabular}{c c c}
				$H$&$E_0$ & $E_2$ \\ 
				22.0&31.82558&$-0.05397$\\
				32.0&31.82558&$-0.11418$\\
				47.0&31.82558&$-0.24632$\\
				59.0&31.82558&$-0.38816$\\
				67.0&31.82558&$-0.50056$\\
				74.0&31.82558&$-0.61061$\\
			\end{tabular}
			\caption{Converted energy scale for TRM protocol at $\Tm=26$ K with $\tw = 10000$ s for various values of $H$ expressed in Oersteds.  The $E_n$ are the magnitude of the nth fitting parameter (including the respective $H^n$) expressed in units of $k_\mathrm{B} T_\mathrm{g}$ (see Eq. \eqref{eq:defEn}).  The correlation length $\xi_\mathrm{TRM}=128(2)$ is derived from Eq.~\eqref{eq:xi_Zeeman_through_teff}. }\label{tabappendixIV}
		\end{ruledtabular}
	\end{centering}
\end{table}

\section{Tables for Sect.~\ref{sec:dependence_Delta_vs_Hd_exp}}\label{appendix:only_tables}

For the reader's convenience we provide in Tables
\ref{tab:zfc375t0}---\ref{tab:trm375} the numerical values used in the
analysis reported in Sect.~\ref{sec:dependence_Delta_vs_Hd_exp}.

\begin{table}[t]
\begin{centering}
\begin{ruledtabular}
\begin{tabular}{c c c  c}
$H$&$E_0$ & $E_2$ & $E_4$\\ 
10.0&33.55848&$-0.06004$&0.00169\\
16.0&33.55848&$-0.15371$&0.01109\\
22.0&33.55848&$-0.29061$&0.03963\\
24.9&33.55848&$-0.37228$&0.06504\\
27.5&33.55848&$-0.45408$&0.09676\\
29.8&33.55848&$-0.53321$&0.13342\\
32.0&33.55848&$-0.61485$&0.17741\\
36.3&33.55848&$-0.79119$&0.29376\\
40.2&33.55848&$-0.97033$&0.44185\\
43.7&33.55848&$-1.14665$&0.61701\\
47.0&33.55848&$-1.32637$&0.82558\\
\end{tabular}
\caption{Converted energy scale for the ZFC protocol at $T=37.5$ K for $\tw = 2500$~s for various values of $H$ expressed in Oersted.  The $E_n$ are the magnitude of the nth fitting parameter (including the respective $H^n$) expressed in units of $k_\mathrm{B} T_\mathrm{g}$ (see Eq. \eqref{eq:defEn}).  The correlation length $\xi_\mathrm{ZFC}=220(20)$ is derived from Eq.~\eqref{eq:xi_Zeeman_through_teff}. }\label{tab:zfc375t0}
\end{ruledtabular}
\end{centering}
\end{table}

\begin{table}[t]
\begin{centering}
\begin{ruledtabular}
\begin{tabular}{c c c c}
$H$&$E_0$ & $E_2$ & $E_4$\\ 
10.0&33.58718&$-0.05392$&0.00043\\
16.0&33.58718&$-0.13804$&0.00284\\
22.0&33.58718&$-0.26099$&0.01014\\
24.9&33.58718&$-0.33433$&0.01664\\
27.5&33.58718&$-0.40780$&0.02475\\
29.8&33.58718&$-0.47886$&0.03413\\
32.0&33.58718&$-0.55218$&0.04538\\
36.3&33.58718&$-0.71055$&0.07515\\
40.2&33.58718&$-0.87143$&0.11303\\
43.7&33.58718&$-1.02978$&0.15784\\
47.0&33.58718&$-1.19117$&0.21120\\
\end{tabular}
\caption{Converted energy scale for the TRM protocol at $\Tm=37.5$ K for $\tw = 2500$~s for various values of $H$ expressed in Oersted.  The $E_n$ are the magnitude of the nth fitting parameter (including the respective $H^n$) expressed in units of $k_\mathrm{B} \Tg$ (see Eq. \eqref{eq:defEn}).  The correlation length $\xi_\mathrm{TRM}=210(16)$ is derived from Eq.~\eqref{eq:xi_Zeeman_through_teff}.}\label{tab:trm375t0}
\end{ruledtabular}
\end{centering}
\end{table}
\begin{table}[t]
\begin{centering}
\begin{ruledtabular}
\begin{tabular}{c c c  c c}
$H$&$E_0$ & $E_2$ & $E_4$ & $E_6$\\ 
10.0&34.11658&$-0.11775$&0.00540&$-0.00011$\\
16.0&34.11658&$-0.30144$&0.03536&$-0.00181$\\
22.0&34.11658&$-0.56990$&0.12639&$-0.01225$\\
24.9&34.11658&$-0.73005$&0.20740&$-0.02575$\\
27.5&34.11658&$-0.89047$&0.30857&$-0.04673$\\
29.8&34.11658&$-1.04565$&0.42548&$-0.07567$\\
32.0&34.11658&$-1.20574$&0.56574&$-0.11602$\\
36.3&34.11658&$-1.55156$&0.93680&$-0.24721$\\
40.2&34.11658&$-1.90286$&1.40904&$-0.45602$\\
43.7&34.11658&$-2.24863$&1.96763&$-0.75252$\\
47.0&34.11658&$-2.60106$&2.63275&$-1.16470$\\
50.3&34.11658&$-2.97914$&3.45374&$-1.74999$\\
56.2&34.11658&$-3.71901$&5.38225&$-3.40443$\\
\end{tabular}
\caption{Converted energy scale for the ZFC protocol at $\Tm=37.5$ K for $\tw = 5000$~s for various values of $H$ expressed in Oersted.  The $E_n$ are the magnitude of the nth fitting parameter (including the respective $H^n$) expressed in units of $k_\mathrm{B} T_\mathrm{g}$ (see Eq. \eqref{eq:defEn}).  The correlation length $\xi_\mathrm{ZFC}=270(20)$ is derived from Eq.~\eqref{eq:xi_Zeeman_through_teff}. }\label{tab:zfc375}
\end{ruledtabular}
\end{centering}
\end{table}
\begin{table}[h]
\begin{centering}
\begin{ruledtabular}
\begin{tabular}{c c c c c}
$H$&$E_0$ & $E_2$ & $E_4$ & $E_6$\\ 
10.0&34.07752&$-0.06364$&0.00034&$-0.00000$\\
16.0&34.07752&$-0.16292$&0.00225&$-0.00000$\\
22.0&34.07752&$-0.30801$&0.00805&$-0.00003$\\
24.9&34.07752&$-0.39457$&0.01322&$-0.00007$\\
27.5&34.07752&$-0.48127$&0.01966&$-0.00012$\\
29.8&34.07752&$-0.56514$&0.02711&$-0.00019$\\
32.0&34.07752&$-0.65166$&0.03605&$-0.00030$\\
36.3&34.07752&$-0.83856$&0.05969&$-0.00063$\\
40.2&34.07752&$-1.02843$&0.08978&$-0.00116$\\
43.7&34.07752&$-1.21530$&0.12538&$-0.00192$\\
47.0&34.07752&$-1.40578$&0.16776&$-0.00297$\\
50.3&34.07752&$-1.61012$&0.22007&$-0.00446$\\
53.3&34.07752&$-1.80791$&0.27746&$-0.00631$\\
56.2&34.07752&$-2.00999$&0.34295&$-0.00867$\\
\end{tabular}
\caption{Converted energy scale for the TRM protocol at $T=37.5$ K for $\tw = 5000$~s for various values of $H$ expressed in Oersted.  The $E_n$ are the magnitude of the nth fitting parameter (including the respective $H^n$) expressed in units of $k_\mathrm{B} T_\mathrm{g}$ (see Eq. \eqref{eq:defEn}).  The correlation length $\xi_\mathrm{TRM}=220(30)$ is derived from Eq.~\eqref{eq:xi_Zeeman_through_teff}.}\label{tab:trm375}
\end{ruledtabular}
\end{centering}
\end{table}

\section{Experimental cooling protocols and numerical simulations}
\label{appendix:cooling_protocols}

The experimental cooling protocols are based on the cooling of the system at a constant speed (e.g. Kelvin degrees per minute) in order to reach the target temperature, usually in the glassy region. In this work we present numerical results based on a sudden or direct quench of the system: i.e. we start at infinite temperature and suddenly we put the system at the target temperature, or in other words, we are implementing an infinite cooling speed. 

One could argue that these two protocols (experimental and direct quench ones)  will produce different behaviors in the measured quantities. In this Appendix we will discuss this issue and we will conclude that one can use the direct or sudden quench in order to study the system and to match with experimental observables obtained with cooling protocols at constant speed.

In Ref.~\cite{janus:14b} we performed a throughout study on the dependence of the dynamics on the cooling protocol. In particular we studied the so-called direct quench and different annealing ones (with different initial temperatures and cooling speeds) for the three-dimensional Edwards-Anderson model in presence of a Gaussian external magnetic field  with variances $H=0.1$ to $0.3$. Let us notice that the model in presence of a Gaussian magnetic field belongs to the same universality class as the one having constant magnetic fields. The main findings of this analysis were the following:
\begin{enumerate}
 \item Despite having different evolutions, the direct quench and the different annealing protocols need the same time to reach the equilibrium, just a bit below the critical temperature in absence of magnetic field (see Fig 3 of Ref.~\cite{janus:14b}). It is important to quote that one of the observables we have studied is just the staggered magnetization of the system, which is the context of this paper corresponds with the field-cooled magnetization.

 \item By fitting  the dynamical behavior of the observables in the glassy region using an stretched exponential we found that the exponent ($\beta$) of this stretched exponential does not depend on the annealing protocol. Moreover, the same happens for the characteristic times of the dynamics (see Table III of 
 Ref.~\cite{janus:14b}).

\item At last, we have performed different works in which we have compared the correlation length obtained in a numerical simulation using a direct quench protocol with those obtained in experiments with a constant speed cooling protocol and the agreement has been very good (see for example Refs.~\cite{janus:17b,zhai-janus:20a,zhai-janus:21}).

 \end{enumerate}

\begin{figure}[h]
	\centering
	\includegraphics[width = 1\columnwidth]{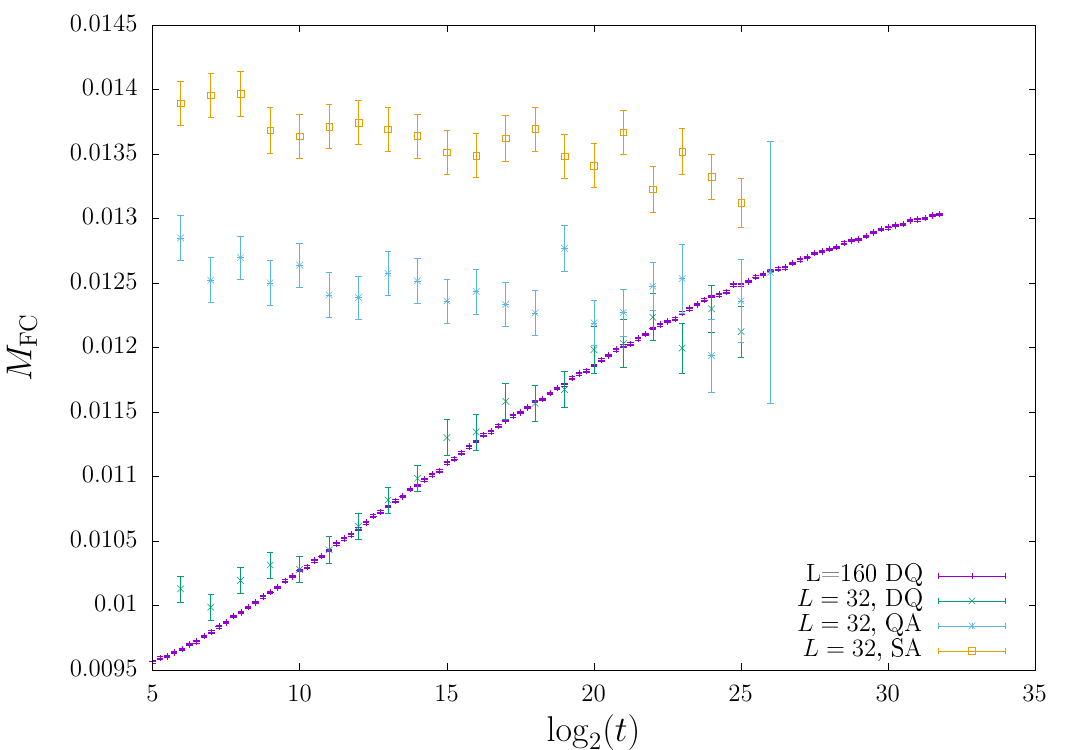}
	\caption{Comparison of the behavior of field-cooling magnetization for a direct quenched (DQ) for a $L=160$ lattice (Janus reference data for this paper, see Fig. \ref{fig:superposition_regime_M_overH}) and a $L=32$ one. In addition we have plotted two different annealing protocols, named slow and quick annealing (SA and QA respectively), both obtained in a $L=32$ lattice. The protocol of the SA was: $T_\text{max}=2.25$, $T_\text{min}=1.0$, $\Delta T=0.25$ and  $2^{16}-2$ sweeps at each temperature different from $T_\mathrm{min}$. The parameters of the QA protocol are the same as in the SA one but we have done $2^{11}-2$ sweeps at each temperature. In all cases $H=0.02$.} 
\label{fig:behavior_fcm}
\end{figure}

Finally, in Fig. \ref{fig:superposition_regime_M_overH} is possible to see that the field-cooled magnetization is growing with time in confront with  the typical experimental behavior of this magnitude, which is essentially constant with time.  The behavior of this observable in numerical simulations is explained by the use of a direct quench: we start the simulation with configuration with zero magnetization, and this magnetization, evolving in a field,  must increase to reach the equilibrium value. 

In Fig. \ref{fig:behavior_fcm} we have analyzed in detail the behavior of the 
field-cooled magnetization from numerical simulations using different annealing protocols as well as the direct quench one. For slow annealing protocols we recover the behavior observed in experiments: the field-cooled magnetization is essentially constant.

\section{The microscopical correlation length}
\label{appendix:xi_micro_def}

Let us define the replicon propagator~\cite{dealmeida:78,dedominicis:06} as:
\begin{equation}\label{eq:Gr_def}
{\mathcal G}_{\text {R}}({\boldsymbol r},t,T)\!=\!\frac {1}{V}\!\sum_{\boldsymbol
x} {\overline {(\langle s_{{\boldsymbol x},t}s_{{\boldsymbol x+ \boldsymbol
r},t}\rangle_T - \langle s_{{\boldsymbol x},t}\rangle_T\langle s_{{\boldsymbol
x+ \boldsymbol r},t}\rangle_T)^2}} \; .
\end{equation}
The replicon correlator ${\mathcal G}_{\text {R}}$ decays to zero in the long-distance limit.  We therefore compute $\xi_{\text {micro}}(\tw;H)$ by exploiting the integral estimators~\cite{janus:08,janus:09}:
\begin{equation}\label{eq:Ik_def}
I_k(t;T)=\int_0^\infty d \, r\,r^k{\mathcal G}({\boldsymbol r}, t;T),
\end{equation}
where
\begin{equation}\label{eq:xi_micro_def}
\xi_{k,k+1}(t,T)={\frac {I_{k+1}(t,T)}{I_k(t,T)}}.
\end{equation}
The $\xi_{12}(\tw;T)$ is designated as the microscopic correlation length $\xi_\mathrm{micro}(\tw;T)$.  

\section{\boldmath Evaluation of the relaxation function $S(t,\tw;H)$ in the TRM protocol}
\label{appendix:relaxation_S}

In this Appendix, we want to convince the reader that the two experimental protocols (TRM/ZFC), which are equivalent in the $H \to 0$ limit, can be treated with the same numerical protocol developed in Refs.~\cite{zhai-janus:20a,zhai-janus:21}.

\begin{figure}[h!]
\centering
\includegraphics[width = 1\columnwidth]{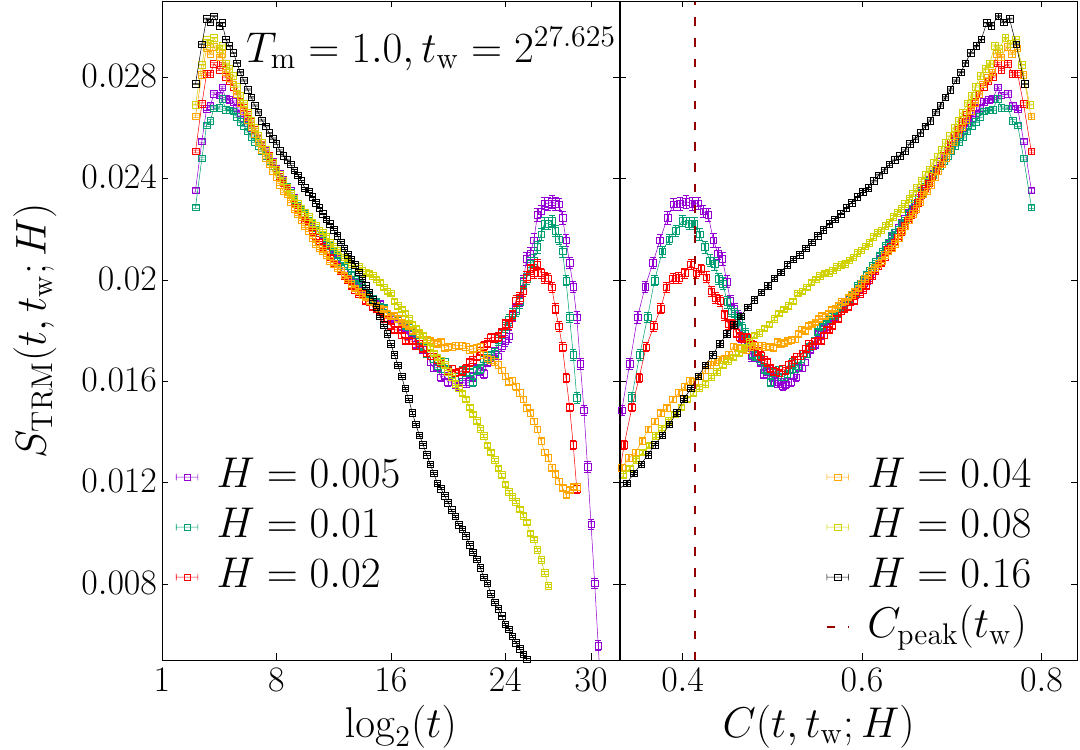}
\caption{A typical set of simulated relaxation functions,
$S_\mathrm{TRM}(t,\tw;H)$.  The data are taken from Run 5, with $\Tm=1.0$ and
$\tw=2^{27.625}$.  \textbf{Left:} 
$S_\mathrm{TRM}(t,\tw;H)$ as a function of time.  \textbf{Right:}
 $S_\mathrm{TRM}(t,\tw;H)$ as a function of the temporal
auto-correlation function $C(t,\tw;H)$.  The dashed line indicates the value of
$C_\mathrm{peak}(\tw)$ (see Table \ref{tab:details_num}).}
\label{fig:relaxation_S_TRM}
\end{figure}

In Fig.~\ref{fig:relaxation_S_TRM}, we exhibit a typical set of relaxation function $S_\mathrm{TRM}(t,\tw;H)$, see Eq.~\eqref{eq:relaxationS_def} for its definition.

One key point unveiled in Refs.~\cite{zhai-janus:20a,zhai-janus:21} was the possibility to define the effective time $\teff_H$ as the time when $C(t,\tw;H)$ reaches the value $C_\mathrm{peak}(\tw)$:
\begin{equation}\label{eq:teff_def_num}
C(\teff_H,\tw;H)=C_\mathrm{peak}(\tw)~.
\end{equation}

\begin{figure}[h!]
	\centering
	\includegraphics[width = 1\columnwidth]{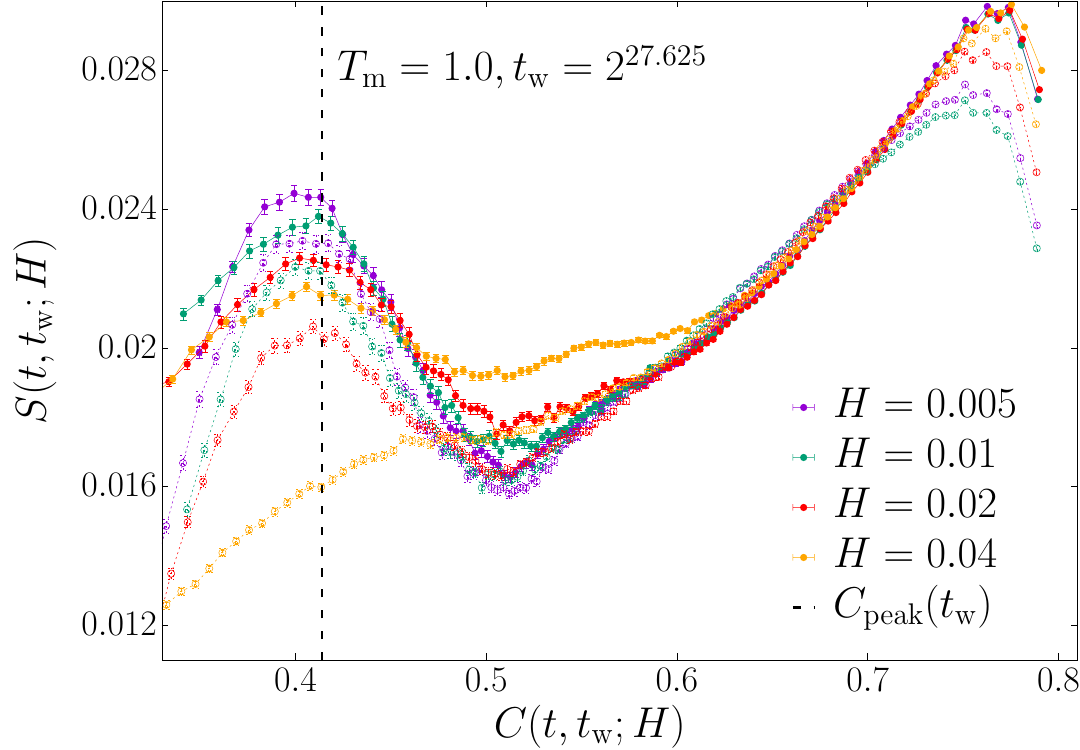}
	\caption{Comparison between the TRM and the ZFC relaxation functions for Run 5, with $\Tm=1.0$ and $\tw=2^{27.625}$.  The empty points are for $S_\mathrm{TRM}(t,\tw;H)$, while the full dots are for the $S_\mathrm{ZFC}(t,\tw;H)$.  The dashed line displays the value for $C_\mathrm{peak}(\tw)$ (see Table \ref{tab:details_num}).  The ZFC points are taken from Refs.~\cite{zhai-janus:20a,zhai-janus:21}.}
	\label{fig:comparison_S_TRM_vs_ZFC}
\end{figure}

As shown in Fig.~\ref{fig:comparison_S_TRM_vs_ZFC}, this physical feature holds
for the TRM protocol as well.

In the following sub-sections, exploiting the behavior of the Hamming distance, we will show how the value of the effective time, $\teff_H$, is independent of the value of $C_\mathrm{peak}(\tw)$ and unveils the physical meaning of Eq.~\eqref{eq:teff_def_num}.

\subsection{Hamming distance: Scaling}
\label{subsec:Hd_connection_to_Ez_num}

We extract the Hamming distance, or at least a surrogate of it, from our knowledge of the
temporal auto-correlation function $C(t,\tw;H)$:
\begin{equation}\label{eq:Hd_def_num}
{\text {Hd}}(t,\tw;H)={\frac {1}{2}}\big[1-C(t,\tw;H)\big]\,.
\end{equation}
A discussion of the connection between the above \emph{numerical} Hamming
distance and the dynamics in the ultrametric tree of states is provided in
Appendix~\ref{appendix:Hd-Cttw}.

\begin{figure}[h]
	\centering
	\includegraphics[width = 1\columnwidth]{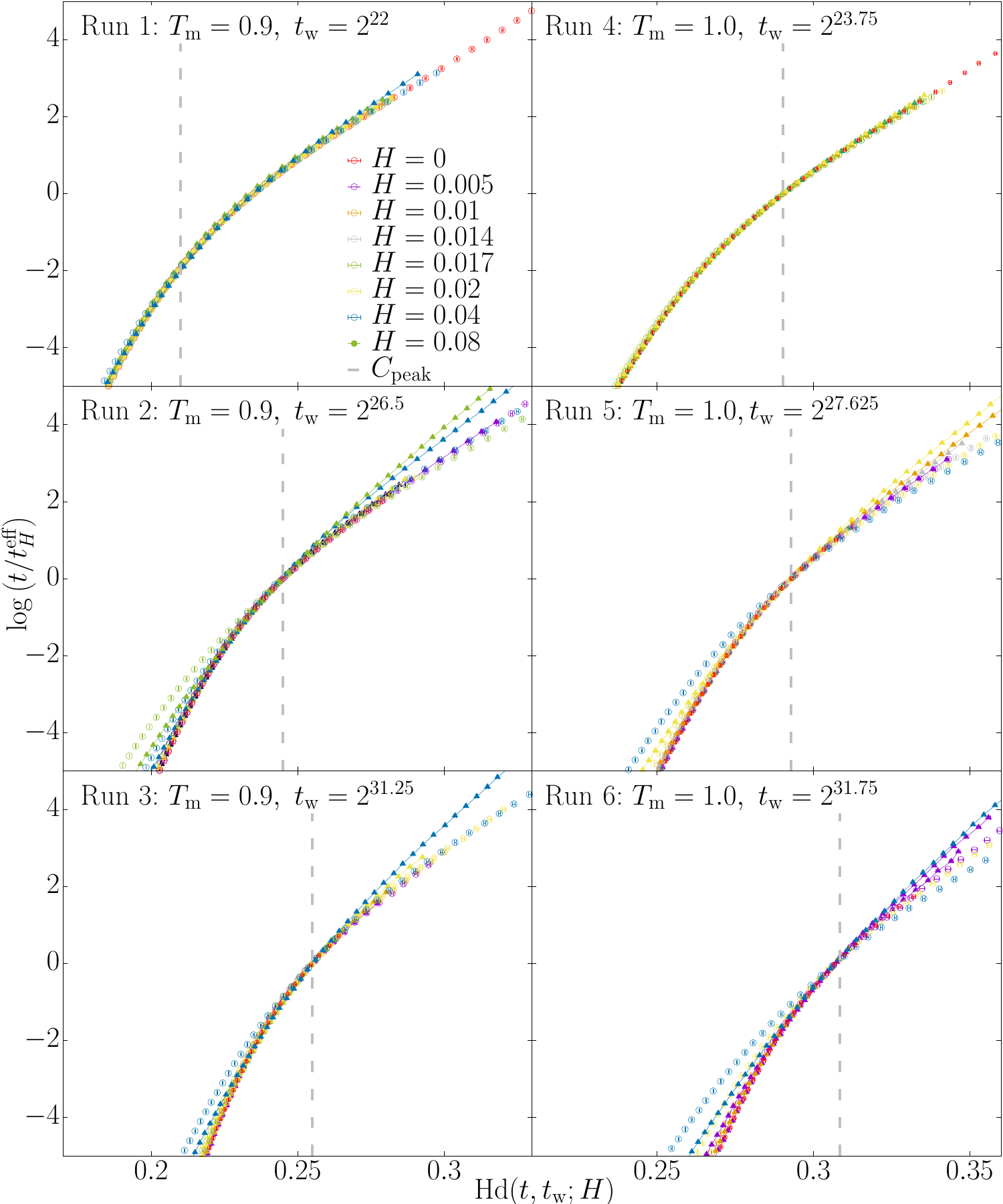}
	\caption{Behavior of the rescaled time $\log(t/\teff_H)$ as a function of the Hamming distance $\mathrm{Hd}(t,\tw;H)$.  The empty circles represent the ZFC data, while the full triangles and joined with lines represent the TRM data.}
	\label{fig:Hd_scaling_num}
\end{figure}

If one displays $\log (t/\teff_H)$ as a function of $\mathrm{Hd}(t,\tw;H)$ at the two simulation temperatures, $T=0.9$ and $T=1.0$, a scaling behavior is apparent from Fig.~\ref{fig:Hd_scaling_num}, with
\begin{equation}\label{eq:Hd_scaling_implication}
\log \big\{t/\teff_H[C_\mathrm{peak}(\tw)]\big\}=\mathcal{F} \big[C(t,\tw;H),\tw\big]~~.
\end{equation}

The determination of the precise value for $C_\mathrm{peak}(\tw)$ is not crucial because $C_\mathrm{peak}(\tw)$ changes $\log \teff_H(C_\mathrm{peak})$ only by a constant.  This implies that $\log \teff_H(C_\mathrm{peak})$ does not depend upon $H^2$.

Developing this important concept further, from Eq.~\eqref{eq:Hd_scaling_implication} we can write,
\begin{equation}\label{eq:H2_independence_of_Cpeak}
\begin{split}
\log \bigg[{\frac {\teff_H(C)}{t_{H}^{\text {eff}}(C_\mathrm{peak})}}\bigg]=\mathcal{F}(C,\tw)=\log\bigg[{\frac {t_{H \to 0^+}^{\text {eff}}(C)}{t_{H \to 0^+}^{\text {eff}}(C_\mathrm{peak})}}\bigg]\\
\Rightarrow \log\bigg[{\frac {\teff_H(C)}{t_{H \to 0^+}^{\text {eff}}(C)}}\bigg]=\log\bigg[{\frac {\teff_H(C_\mathrm{peak})}{t_{H \to 0^+}^{\text {eff}}(C_\mathrm{peak})}}\bigg]~~,
\end{split}
\end{equation}
implying that the value of the effective time, $\teff_H$, is {\it independent} of the value of $C_\mathrm{peak}(\tw)$.

\subsection{\boldmath Extraction of the effective response time $\teff_H$}
\label{subsec:teff_extraction_num}
In this Appendix, we extract the effective response time $\teff_H$ for the TRM cases, and we demonstrate the validity of the scaling law in Eq.~\eqref{eq:scaling_law_decay_teff} for the TRM case as well.  We show the decay of $\teff_H$ as a function of $H^2$ in Fig.~\ref{fig:log_teff_vs_H2_num}, along with those for the ZFC protocol (see below).  The data for $\log (\teff_H/ \teff_{H \to 0^+})$ are fitted by the function,
\begin{equation}\label{eq:fitting_teff_num}
f(x)=a_2(\tw;T)x+{\mathcal {O}}(x^2)~~,
\end{equation}
where $x=H^2$.  Remember that $a_0=\log \teff_{H \to 0^+}$. In order to avoid the unphysical wild oscillations at large
magnetic fields (recall that $H=1$ for the IEA model roughly corresponds to
$5\times 10^4$~Oe in physical units~\cite{zhai-janus:20a}), we define a unique
fitting range in the small $x$ region, $x=H^2\in [0,0.0003]$.  Our fitting
parameters are displayed in Table \ref{tab:ZFC_fit_num} for the ZFC data, and
Table \ref{tab:TRM_fit_num} for the TRM data.

\begin{figure}[h]
	\centering
	\includegraphics[width = 1\columnwidth]{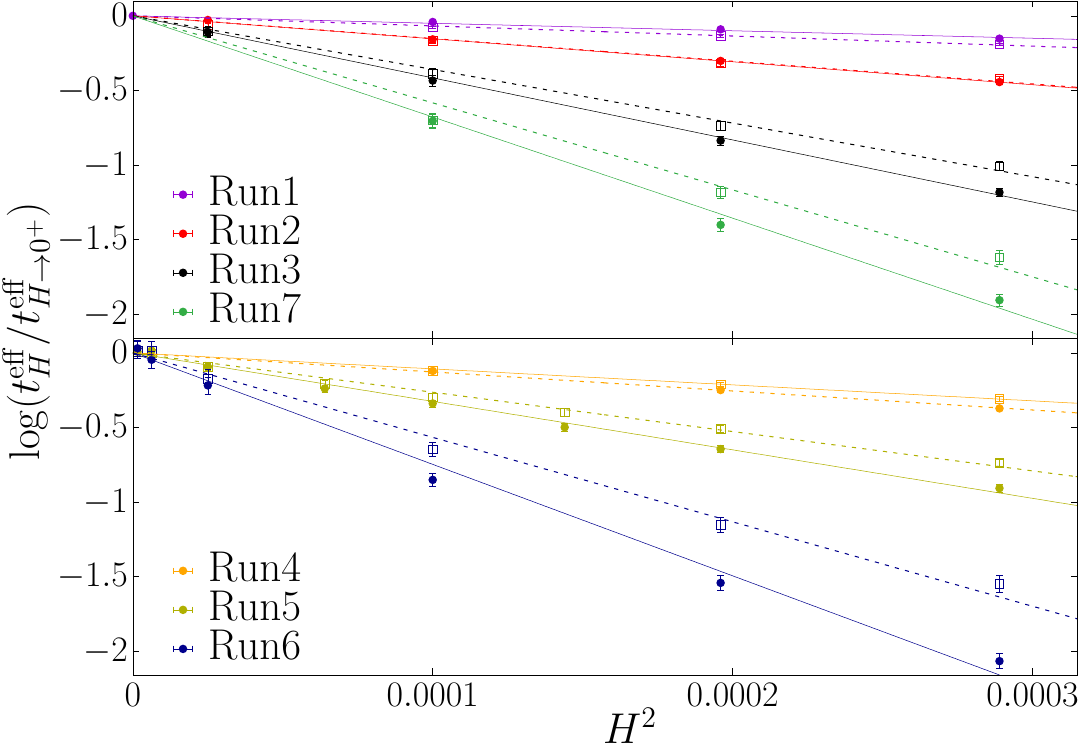}
	\caption{The numerical ratio of $\log(\teff_H/\teff_{H\rightarrow 0^+})$ for the seven runs defined in Table \ref{tab:details_num} for both the ZFC and TRM protocols.  The filled dots refer to the ZFC protocol; the empty squares to the TRM protocol.  The coefficients of the $H^2$ fit, $a_2(\tw;T)$ from Eq.~\eqref{eq:fitting_teff_num} are listed in Table \ref{tab:ZFC_fit_num} for the ZFC data, and Table \ref{tab:TRM_fit_num} for the TRM data.  The continuous lines represent the fit to the ZFC data, while the dashed lines represent the fit to the TRM data.  The ZFC data are the same as in Refs.~\cite{zhai-janus:20a,zhai-janus:21}.}
	\label{fig:log_teff_vs_H2_num}
\end{figure}
\begin{figure}[h]
	\centering
	\includegraphics[width = 1\columnwidth]{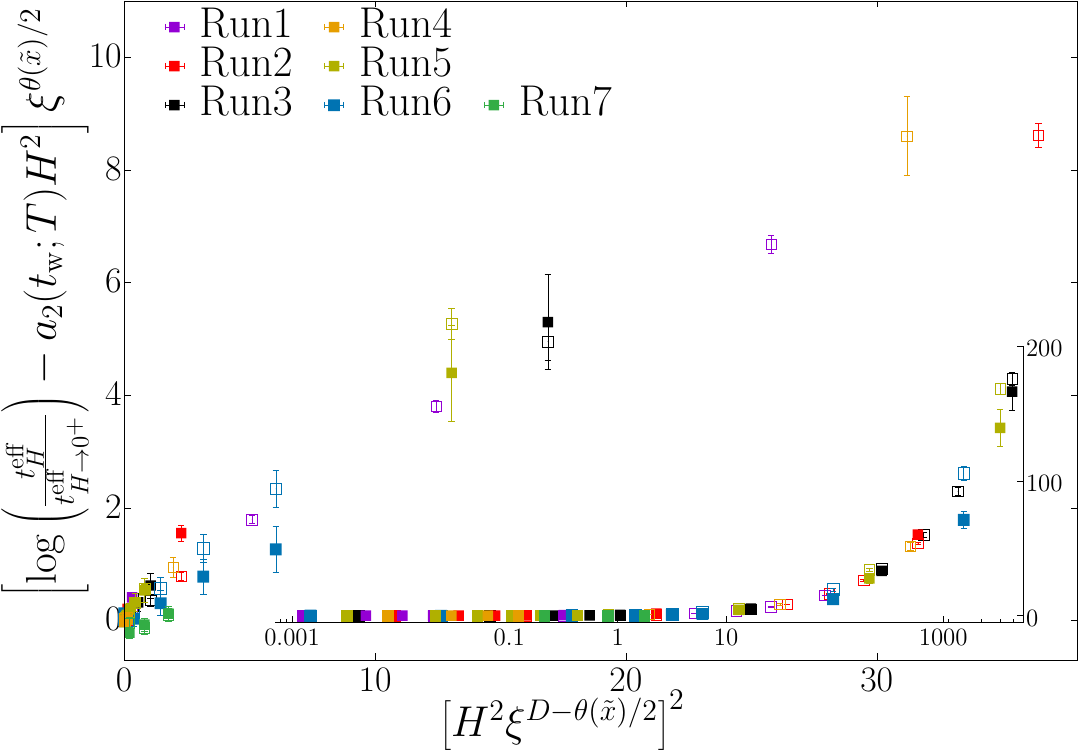}
	\caption{The non-linear parts from the numerical response time data, $[\log (\teff_H/\teff_{H\rightarrow 0^+})-c_2(\tw,T)H^2]\xi^{\theta({\tilde  x})/2}$ plotted against $(\xi^{D-\theta({\tilde x})/2}H^2)^2$.  The abscissa of the {\it main panel} is in a linear scale, showing an expanded view for small values of $(\xi^{D-\theta({\tilde x})/2}H^2)^2$.  The abscissa of the {\it insert} is in a log scale in order to report all of our numerical data.  The open squares refer to the ZFC data, while the filled squares refer to the TRM data.}
\label{fig:scaling_law_num}
\end{figure}

\begin{table}[!h]
\begin{centering}
\begin{ruledtabular}
\begin{tabular}{c c c c}
$T$&$\tw$ &Coefficient & Numerical value\\
\hline\\[-5pt]
0.9&$2^{22}$&$a_2$&$-5.01(14)\times 10^2$\\
0.9&$2^{26.5}$&$a_2$&$-1.54(2)\times 10^3$\\
0.9&$2^{31.25}$&$a_2$&$-4.13(11)\times 10^3$\\
0.9&$2^{34}$&$a_2$&$-6.78(13)\times 10^3$\\
&&&\\
1.0&$2^{23.75}$&$a_2$&$-1.29(2)\times 10^3$\\
1.0&$2^{27.625}$&$a_2$&$-3.25(3)\times 10^3$\\
1.0&$2^{31.75}$&$a_2$&$-7.48(17)\times 10^3$
\end{tabular}
\caption{Results for the fit to Eq.~\eqref{eq:fitting_teff_num} for the ZFC data for the time ratio $\log(\teff_H/\teff_{H\rightarrow 0^+})$.  The fitting range is $0\leq H^2\leq0.0003$.}
\label{tab:ZFC_fit_num}
\end{ruledtabular}
\end{centering}
\end{table}
\begin{table}[!h]
\begin{centering}
\begin{ruledtabular}
\begin{tabular}{c c c c}
$T$&$\tw$ &Coefficient & Numerical value\\ 
\hline\\[-5pt]
0.9&$2^{22}$&$a_2$&$-6.77(11)\times 10^2$\\
0.9&$2^{26.5}$&$a_2$&$-1.52(2)\times 10^3$\\
0.9&$2^{31.25}$&$a_2$&$-3.60(14)\times 10^3$\\
0.9&$2^{34}$&$a_2$&$-5.84(16)\times 10^3$\\
&&&\\
1.0&$2^{23.75}$&$a_2$&$-1.06(1)\times 10^3$\\
1.0&$2^{27.625}$&$a_2$&$-2.64(3)\times 10^3$\\
1.0&$2^{31.75}$&$a_2$&$-5.65(22)\times 10^3$
\end{tabular}
\caption{Results for the fit to Eq.~\eqref{eq:fitting_teff_num} for the TRM data for the time ratio $\log(\teff_H/\teff_{H\rightarrow 0^+})$.  The fitting range is $0\leq H^2\leq0.0003$.}
\label{tab:TRM_fit_num}
\end{ruledtabular}
\end{centering}
\end{table}
Finally,  in Fig.~\ref{fig:scaling_law_num} we show that the scaling law in Eq.~\eqref{eq:scaling_law_decay_teff} holds for both the ZFC and TRM protocols.

%\newpage
\section{A discussion of Eq.~\eqref{eq:Hd_def_num}}\label{appendix:Hd-Cttw}

In a similar vein to Appendix~\ref{appendix:phenomenological_HD}, we
connect here the dynamics in the abstract ultrametric tree of states
with the operational definition of the Hamming distance that we
provided in Eq.~\eqref{eq:Hd_def_num}. We rely on an exponential
increase in the degeneracy of states with decreasing overlap
$q_{\alpha\beta}$ from their respective initial state as a consequence
of an underlying ultrametric topology of overlap 
space~\cite{mezard:84}.

An initial state (at $\tw=0$) will evolve into new states as time progresses.
It is useful to define a
distance, the Hamming distance, in terms of the overlap of states as time
progresses. Consider three states $\alpha,~\beta,$ and
$\gamma$ and their overlaps $q_{\alpha\beta},~ q_{\alpha\gamma}$ and
$q_{\beta\gamma}$~\cite{mezard:84}.  Order them $q_{\alpha\beta}\geq q_{\alpha\gamma} \geq
q_{\beta\gamma}$.  The ultrametric topology of the pure states, which we assign
to metastable states as well~\citep{hammann:92}, results in
$q_{\alpha\beta}\geq q_{\alpha\gamma}=q_{\beta\gamma}$.

Figure ~\ref{fig:tree_daughter_random} is a pictorial representation of this
relationship.  A tree is constructed in overlap space, with the level at
$\Tm$ representing the pure (metastable) states of the system at $\Tm$.  The
states are grouped in such a way that the ultrametric topology is represented
by the ``branches'' that connect the states. 
Consider the first state (to the left) of the bottom level.  Its
overlap with itself, $q_{\alpha\alpha}\equiv q_{\text {EA}}$, and its value is
a function of temperature as illustrated in
Fig.~\ref{fig:tree_daughter_random}.  The next state, $\beta$, is connected to
$\alpha$ by a branch to the first level above the bottom, diminishing $q$ and so on.
We represent the distance along the level at $\Tm$ by the Hamming
distance, $\mathrm{Hd} = (1/2)(q_{\alpha\alpha}-q_{\alpha\phi})\equiv
(1/2)(q_{\text {EA}}-q_{\alpha\phi})$ where $\phi$ represents some state
further along the bottom level.

We interpret the time evolution of the metastable states in terms of diffusion along a level of the tree from the initial state $\alpha$ to states with every diminishing overlap.  The ultrametric geometry of the state space leads to an exponential increase in the number (degeneracy) of states encountered in time $\tw+t$, as the overlap $q_{\alpha\phi}$ diminishes.  We associate a barrier 
height $\Delta_{\alpha\phi}$ proportional to a function of the reduction in overlap between states $\alpha$ and $\phi$, and hence to a function of $\mathrm{Hd}$.  
Thus, at finite temperatures and times, $\tw+t$, there will be a maximum barrier overcome by the spin-glass system associated with a minimum overlap $q_{\text {min}}$ set by the time $t+\tw$. This is borne out experimentally (see Ref.~\citep{hammann:92}).

Thus, as time progresses, because of the exponential increase in the number of
states associated with minimum overlap $q_{\alpha\phi}=q_{\text {min}}$,
essentially all of the decrease in the occupation of the initial state can be
found in the occupation of the states lying at a Hamming distance associated
with states of minimum overlap with state $\alpha$.  This allows us to write Eq.~\eqref{eq:Hd_def_num}.

To summarize, the Hamming distance, Eq.~\eqref{eq:Hd_def_num}, expresses the transfer of the population of the states of the spin glass at $t=0$, the time of the temperature quench to $\Tm$, to the population of states with minimum overlap with those states at $t = \tw$ when the magnetic field $H$ is turned off (TRM) or on (ZFC).

\section{Scaling behavior of the Hamming distance}\label{appendix:scaling}

In this section we will re-derive, in the framework of the renormalization group~\cite{amit:05}, the dependence of the Hamming distance with $\xi$.

Since we are evolving at the minimum available overlap (denoted as $q_\mathrm{min}$), the dynamics is controlled by the replicon mode~~\cite{dedominicis:98}, which has the following correlation function behaving, for large $x$, as~\cite{dedominicis:98}
\begin{equation}
\label{eq:replicon}
\langle q(x) q(0) \rangle_{q=q_\mathrm{min}}  \sim x^{-\theta}\,,
\end{equation}
which defines the replicon exponent $\theta$, already quoted in the text, and the associated dimension (momentum) of the field $q(x)$, $\mathrm{dim}(q)=\theta/2$.~\cite{amit:05}

If we turn on a magnetic field, an additional term appears in the Hamiltonian~\cite{bray:80}, that can be written in the continuum as
\begin{equation}
\label{eq:extraS}
+ H^2 \int d^D x ~q(x)\,.
\end{equation}
Using this equation, we can write the following relation between the 
anomalous dimensions (in momentum) of the observables $H^2$ and $q(x)$ (in the field theoretic approach the Hamiltonian is dimensionless):
\begin{equation}
L^0 = L^{-\mathrm{dim}(H^2)+D -\mathrm{dim}(q)}\,,
\end{equation}
and then
\begin{equation}
\label{eq:extraS2}
D-\mathrm{dim}(H^2) -\mathrm{dim}(q)=0\,,
\end{equation}
where, by Eq. (\ref{eq:replicon}), 
\begin{equation}
\label{eq:dimH2}
\mathrm{dim}(H^2) =D-\theta/2\,.
\end{equation}

The Zeeman energy is given by Eq.~(\ref{eq:extraS})
\begin{equation}
E_\mathrm{Z}=H^2 \int d^D x ~q(x) \equiv N_\mathrm{eff} H^2\,,
\end{equation}
where $N_\mathrm{eff}= \int d^Dx~q(x)$ and it will evolve with $\xi$, accordingly with their dimensions, as
\begin{equation}
N_\mathrm{eff} \sim \xi^{D-\mathrm{dim}(q)}
\sim \xi^{\mathrm{dim}(H^2)} 
\sim \xi^{D-\theta/2}\,.
\end{equation}

The Hamming distance, which is proportional to $q$, will evolve with the dynamical correlation length scale as
\begin{equation}
L^D \mathrm{Hd} \sim \xi^{(D-{\mathrm{dim}(q)})}\sim \xi^{(D-\theta/2)}\sim N_\mathrm{eff} \,,
\end{equation}
which is the behavior of $N \mathrm{Hd}$ derived in Sec. \ref{sec:dependence_Delta_vs_Hd_exp}
 using another approach ($N=L^D$ being the number of spins of the system).

\bibliographystyle{apsrev-4_1-titles}
\bibliography{PRB_final.bbl}

\end{document}